\pdfoutput=1
\documentclass{aa}
\usepackage{graphicx}
\usepackage{txfonts}
\usepackage{subfigure}
\newcommand{\kms}{km\,s$^{-1}$}
\newcommand{\ms}{m\,s$^{-1}$}

\begin{document}

    \title{Anti-solar differential rotation on the active sub-giant\\ HU~Virginis \thanks{Based on data obtained with the STELLA robotic observatory in Tenerife, an AIP facility jointly operated by AIP and IAC.}$^{,}$\thanks{The Tables with the STELLA radial velocity data set and the APT photometric data set are available only in electronic form at the CDS via anonymous ftp to CDS via anonymous ftp to cdsarc.u-strasbg.fr (130.79.128.5)
or via http://cdsweb.u-strasbg.fr/cgi-bin/qcat?J/A+A/.} }
   \author{G. Harutyunyan, K. G. Strassmeier, A. K\"{u}nstler, T. A. Carroll, \and M. Weber}

   \institute{Leibniz Institute for Astrophysics Potsdam (AIP),
              An der Sternwarte 16, 14482 Potsdam, Germany\\
   \email{[gharutyunyan;kstrassmeier;akuenstler;tcarroll;mweber]@aip.de}
            }
   \date{Received 05 April 2016 $/$ Accepted 18 May 2016}

  \abstract
{Measuring surface differential rotation (DR) on different types of stars is important when characterizing the underlying stellar dynamo. It has been suggested that anti-solar DR laws can occur when strong meridional flows exist. }
{We aim to investigate the differential surface rotation on the primary star of the RS~CVn binary, HU~Vir, by tracking its starspot distribution as a function of time. We also aim to recompute and update the values for several system parameters of the triple system HU~Vir (close and wide orbits). }
{Time-series high-resolution spectroscopy for four continuous months was obtained with the 1.2-m robotic STELLA telescope. Nine consecutive Doppler images were reconstructed from these data, using our line-profile inversion code \emph{iMap}. An image cross-correlation method was applied to derive the surface differential-rotation law for HU~Vir. New orbital elements for the close and the wide orbits were computed using our new STELLA radial velocities (RVs) combined with the RV data available in the literature. Photometric observations were performed with the \emph{Amadeus} Automatic Photoelectric Telescope (APT), providing  contemporaneous Johnson-Cousins $V$ and $I$ data for approximately 20 years. This data was used to determine the stellar rotation period and the active longitudes.}
{We confirm anti-solar DR with a surface shear parameter $\alpha$ of $-0.029 \pm 0.005$ and $-0.026 \pm 0.009,$ using single-term and double-term differential rotation laws, respectively. These values are in good agreement with previously claimed results. The best fit is achieved assuming a solar-like double-term law with a lap time of $\approx$ 400\,d. Our orbital solutions result in a period of 10.387678 $\pm$ 0.000003 days for the close orbit and 2726 $\pm$ 7\,d ($\approx$ 7.5\,yr) for the wide orbit. A Lomb-Scarge (L-S) periodogram of the pre-whitened $V$-band data reveals a strong single peak providing a rotation period of 10.391 $\pm$ 0.008\,d, well synchronized to the short orbit.}
{}


\titlerunning{Anti-solar differential rotation on HU\,Vir}
\authorrunning{G. Harutyunyan et al.}
\maketitle


\section{Introduction}

Differential rotation (DR) is among the key mechanisms for the dynamo action of stars with convective envelopes (see, e.g., \citealt{2013mpap.book.....R}, and references therein). Fast rotation and its related internal momentum distribution, together with turbulent convection, may sustain an internal magnetic field that eventually makes its fingerprint on the stellar surface in form of starspots (e.g., \citealt{2009A&ARv..17..251S}). In close binaries tidal effects help to maintain the fast rotation, as well as magnetic activity at high levels (e.g., \citealt{2000AN....321..175H}, \citeyear{2002AN....323..399H}). Numerical simulations, as well as mean-field magneto-hydrodynamic (MHD) computations, deal with the many aspects of Reynolds stresses and horizontal heat flow in rotating convection, which are responsible for deviation from the cylinder-shaped velocity pattern and thus result in differential rotation (e.g., \citealt{1995A&A...299..446K}, \citeyear{2004AN....325..496K}; \citealt{2011AN....332..933K}, \citeyear{2012AN....333.1028K}, \citeyear{2014IAUS..302..194K}; \citealt{2011A&A...531A.162K}, \citealt{2013ApJ...777L..37W}, \citealt{2014MNRAS.438L..76G}). These simulations even predicted anti-solar DR for stars with fast meridional circulation \citep{2004AN....325..496K}. Despite the recent success of the confirmation of the (effective) temperature dependency of DR for a large number of \emph{Kepler} stars ({\citealt{2013A&A...560A...4R}, \citealt{2015A&A...576A..15R}, \citealt{2015csss...18..535K}), the number of tuneable parameters is still too large  to arrive at a single theory for DR throughout the HR-diagram. Spatially resolved stellar disk observations and subsequent tracking of starspots in time may help to further constrain the theory.

HU\,Virginis (\object{HD\,106225}, K0\,IV, $v\sin i$ = 25\,\kms ) is a rapidly rotating sub-giant star in a close binary system with an unseen secondary component and with an orbital and rotational period of $P_{\rm rot} \approx P_{\rm orb} \approx 10.4$~d. This shows RS\,CVn type activity, such as light and color variability  \citep{1986ApJS...60..551F}, strong \ion{Ca}{II} H\&K emission \citep{1990ApJS...72..191S}, H$\alpha$ and ultraviolet emission \citep{1986ApJS...60..551F}, coronal X-ray emission \citep{1993ApJS...86..599D}, radio emission \citep{1989ApJS...71..905D}, and optical and ultra-violet spectral line variations \citep{1986ApJS...60..551F}.
\cite{1999A&AS..137..369F} found HU\,Vir to be a member of a spectroscopic triple
system with an orbital period of about six years.

\cite{1994A&A...281..395S} derived the first Doppler images of HU\,Vir with the Coud\'e Feed Telescope at the Kitt Peak National Observatory (KPNO) for two epochs in 1991 using four photospheric lines. The derived differential rotation was a factor of 10 less extreme than that on the Sun but, surprisingly, of opposite sign ($\alpha=-0.0228$, converted to have the same definition of $\alpha$ in this paper), suggesting faster rotation of polar regions compared to the equatorial regions. The two images showed a cool polar spot, while H$\alpha$ and \ion{Ca}{II} line-profile analysis revealed two chromospheric hot plages $180^{\circ}$ apart, which seemed to be spatially related to two large appendages of the polar spot. Variable broadening of the H$\alpha$ emission profile suggested mass flow within a closed-loop geometry connecting the two plage regions across the visible pole. Two ROSAT-HRI observations of HU\,Vir in 1994 and 1995 revealed a large and long duration flare lasting for more than 1.5 days but without rotational modulation and thus probably related to the polar loop geometry \citep{1997A&A...328..565E}.

\cite{1998A&A...330..541H} presented an independent Doppler image of HU\,Vir from data obtained with the 2.1-m telescope at McDonald Observatory in 1995, again using four photospheric lines. A large spot at latitude $\approx$ 45$^{\circ}$ and a weak polar spot with one appendage towards lower latitudes were reconstructed, the darkest regions having temperatures 650-870 K below the photospheric temperature. The comparison of the Doppler images for epoch 1995.14 to the one obtained by \cite{1994A&A...281..395S} for epoch 1991.3 confirmed the anti-solar differential rotation of HU\,Vir but rested on the assumption of correct spot identification from images separated in time by nearly four years.

$UBV$ photometry was used to trace the evolution of spotted regions in 1991, and a fairly stable spot distribution was found. Long-term photometric monitoring presented by \cite{2000A&A...356..643O}  suggest the existence of a 5.6-yr variability cycle. Photometric spot modeling of HU\,Vir was carried out by \cite{2002A&A...381..517A} using independent data collected at the South African Astronomical Observatory (SAAO) in 1997. The optical and infrared color curves were modeled simultaneously, obtaining spot temperatures of 4 450 and 4 050\,K for the two reconstructed spots.

In this paper, we present nine consecutive Doppler images of HU\,Vir, based on a continuous four-month long time series of spectroscopic data between February and June 2013. Our new image reconstructions are based on simultaneous inversions from several dozens of spectral lines (multi-line inversions). Section~\ref{S2} introduces the new data that are first used in Sect.~\ref{S3} to refine the system parameters of HU\,Vir prior to further analysis. Sect.~\ref{S4} describes the line-profile inversions and  Sect.~\ref{S5} analyzes and discusses the differential surface rotation. Section~\ref{S6} presents our summary and conclusions.


\section{Observations and data reduction}\label{S2}

\subsection{Spectroscopic observations}

Observations were carried out with the STELLA echelle spectrograph (SES) at the robotic 1.2-m STELLA-I telescope at the Observatorio del Teide in Tenerife, Spain \citep{2010AdAst2010E..19S}. A time series of 118 \'echelle spectra was taken in the period between Feb 17 to Jun 26, 2013. The integration time was set to 3 600\,s for the first 34 spectra, and to 5 400\,s for the following 84 spectra, providing signal-to-noise ratios (S/N) between 40 and 160, depending on the weather conditions. The spectra cover the wavelength range from 386.5\,nm to 882.5\,nm with a resolving power of $R$=55,000, corresponding to a spectral resolution of 0.012\,nm (5.5\,\kms) at 650\,nm.

SES spectra are automatically reduced and extracted using the IRAF-based STELLA data-reduction pipeline \citep[see][]{2008SPIE.7019E..0LW}. The CCD images were corrected for bad pixels and cosmic rays. Bias levels were removed by subtracting the average overscan from each image followed by the subtraction of the mean of the (already overscan subtracted) master bias frame. The target spectra were flattened through a division by  a long-term averaged flat-field spectrum (master flat) which had been normalized to unity.

Radial velocities (RV) from STELLA-SES spectra were derived from an order-by-order cross correlation with a synthetic template spectrum and then averaged. A total of 60 orders out of the 80 available were used and a synthetic spectrum of a K0 subgiant was adopted (for more details, see \citealt{2012AN....333..663S}). The time sequence of our new HU\,Vir RVs are plotted in Fig.~\ref{F1}a. The standard (external) error of a STELLA-SES observation of HU\,Vir is $\approx$ 30\,\ms. A systematic zero point shift of 0.503\,\kms  was added to the data in this paper, see \cite{2012AN....333..663S} for its determination with respect to the CORAVEL system. The complete RV data set is available in electronic form at CDS Strasbourg.

\subsection{Photometric observations}

Photometric observations have been carried out with the {\sl Amadeus} Automatic Photoelectric Telescope (APT) at Fairborn Observatory in southern Arizona \citep{1997PASP..109..697S} for approximately 20 years now starting in 1996. The photometric data points that have errors larger than 20~mmag, as well as the outliers of each observational season, were excluded for the investigations in this paper, resulting in 2 816 and 2 483 Johnson-Cousins $V$ and $I$ data points, respectively, 33 and 17 of which were made during our spectroscopic observations in the first semester of 2013.
The entirety of our {\sl Amadeus} HU\,Vir photometry is shown in Fig.~\ref{F2}a. Measurements were always made differentially with respect to \object{HD 106270} and \object{HD 106332} as a comparison and check star, respectively. For further details of APT data and its reduction we refer to \cite{2001AN....322..325G}. We present the complete photometric data set in electronic form at CDS Strasbourg.

\begin{figure}[!htb]

 {\bf  a)}
 
    \includegraphics[trim= 8 200 40 15,clip,width=87mm]{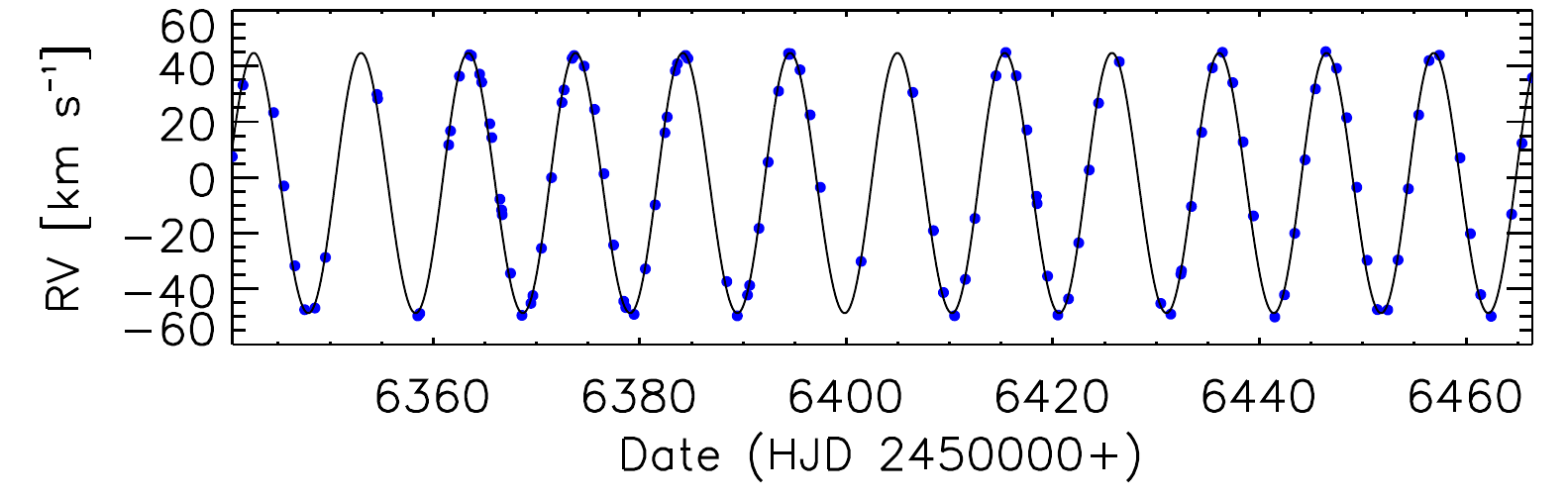}
 {\bf  b)}
 
    \includegraphics[trim= 8 0 40 156,clip,width=87mm]{F1.pdf}

  \caption{STELLA radial velocities of HU\,Vir. $a.$ The entire time series obtained during February through June 2013. $b.$ The phased RV curve before (left) and after spot-jitter removal (right). The respective lower panels show the O--C residuals. Phases were computed from Eq.~\ref{eq:phase}. The solid line is our final orbit from Table~\ref{T1}.}
 \label{F1}
\end{figure}

\section{A brief recap of the astrophysical parameters of HU\,Vir}\label{S3}

\subsection{Orbital elements}\label{S3-1}

The extraction of the orbital elements for a spotted star is complicated by the star's significant line-profile variability, and by the discovery that the system is triple \citep{1999A&AS..137..369F}. A wide orbit of an unseen third component with a period of about 6.1~years was identified by \cite{1999A&AS..137..369F} from a period search done on the residuals of the visible component of the close pair. Here we present a reassessment of the wide and the close orbits by adding our new STELLA data. The technical details of our orbit computation are the same as discussed in \cite{2011A&A...531A..89W} and \cite{2013A&A...559A..17S} and we refer the reader to these papers.

For the initial orbital solution, a total of 118 STELLA RVs were available. Fig.~\ref{F1}a shows the new STELLA RV data as a function of time. Figure~\ref{F1}b (left panel) shows the phased RV curve before spot-jitter removal and the residuals of the initial orbital fit. The standard error of an observation of unit weight was almost 1~\kms , three times the typical RV error for STELLA, and appeared rather systematic and obviously were due to star-spot jitter. We first examined the residuals for periodicity and found well-defined peaks at the rotation period and its first two harmonics. Standard discrete Fourier transforms of the residuals were calculated, followed by a non-linear least-square fit of the frequencies found. The fit was then removed from the data. Fig.~\ref{F1}b (right panel) shows the corrected STELLA RV data and the orbital fit after removal of the star-spot jitter. The achieved standard error of an observation of unit weight was then $\approx$ 200~\ms , thus much closer to the SES instrumental error of $\approx 30$~\ms\ achieved for narrow-lined standard stars \citep{2012AN....333..663S}. This altered the elements with respect to the initial orbit on average  by just a few \%\ but lowered the uncertainties by up to a factor five. A similar analysis had been done for the highly-eccentric active binary \object{HD\,123351} \citep{2011A&A...535A..98S} with a similar reduction of errors. As in our initial solution in \cite{1999A&AS..137..369F}, the short-period eccentricity came out quite small, $0.019$, but nevertheless highly significant and has been retained because in triple systems the third star may cause a non-zero eccentricity (e.g., \citealt{1979A&A....77..145M}) in the short-period orbit.

We then combined the STELLA data with the RV data from \cite{1999A&AS..137..369F} and \cite{1994A&A...281..395S} and recomputed the long-period orbit, assuming the orbital elements from \citeauthor{1999A&AS..137..369F} as starting values. The time baseline was thereby almost doubled. The STELLA RVs were shifted to the same IAU RV system defined by \cite{1990PDAO...18...21S}, as described in \cite{1999A&AS..137..369F}. The initial solution converged on a very similar eccentricity than the \citeauthor{1999A&AS..137..369F} orbit, $0.53 \pm 0.02$, with residuals of 1.75~\kms\ but with a period longer by $\approx$ 500~d. Moreover, a forced $e=0$ solution converges at basically the same residuals and with a comparable period only shorter by 40~d, i.e., 5$\sigma$. Therefore, we adopt the zero-eccentricity orbit in this paper but state that its solution for HU\,Vir remains preliminary. Table~\ref{T1} lists the revised HU\,Vir elements for the close and the wide orbits.

\begin{table}
\begin{flushleft}
\caption{Revised spectroscopic orbital elements.}\label{T1}
\begin{tabular}{lll}
\hline\hline \noalign{\smallskip}
Element & Close orbit & Wide orbit \\
\noalign{\smallskip} \hline \noalign{\smallskip}
$P_{\rm orb}$ (days)           & 10.387678$\pm$0.000003  & 2726$\pm$7 \\
$T_{\rm Periastron}$ (245+)    & 6,337.96$\pm$0.05        & \dots \\
$\gamma$ (km~s$^{-1}$)         & --1.63$\pm$0.02         & \dots \\
$K_{1}$ (km~s$^{-1}$)          & 46.71$\pm$0.03          & 1.92$\pm$0.05\\
$e$                            & 0.0188$\pm$0.0006       & 0 (fixed)\\
$\omega$                       & 191$\pm$2               & \dots\\
$a_1\sin i$ (10$^6$ km)        & 6.6709$\pm$0.0045       & 71.9$\pm$1.8\\
$f(m)$ (M$_{\sun}$)            & 0.1099$\pm$0.0002       & 0.0020$\pm$0.0002\\
$N$                            & 118                     & 289 \\
fit rms (\ms )                 & 213                     & 1750 \\
\noalign{\smallskip} \hline
\end{tabular}
\tablefoot{The close orbit is from the STELLA data with the starspot-jitter removed; the wide orbit is computed combining all literature data.}
\end{flushleft}
\end{table}

We note that the times of the Doppler-imaging observations in this paper, as well as the photometry, are phased with the following ephemeris based on the very precise orbital period:
\begin{equation} \label{eq:phase}
HJD = 2,456,332.4504 + 10.387678 \times\ E \ ,
\end{equation}
where the zero point is adopted as being a time of positive radial velocity.

We note that the mass function, together with our inability to see a secondary star in optical spectra, suggests that the orbit has an appreciable inclination with respect to the plane of the sky. If we assume at least a 2\fm5 brightness difference between the secondary and the primary at red and blue wavelengths, the secondary cannot be of earlier spectral type than G2(V). Adopting the secondary mass as $M_2$ = 1.0 M$_\sun$, the observed $f(m)\approx 0.1$ provides a lower limit to the inclination of $\approx$ 50\degr.

 \begin{figure}[!htb]
 {\bf  a)}

 \includegraphics[angle=0,trim= 0 0 0 15, width=87mm,clip]{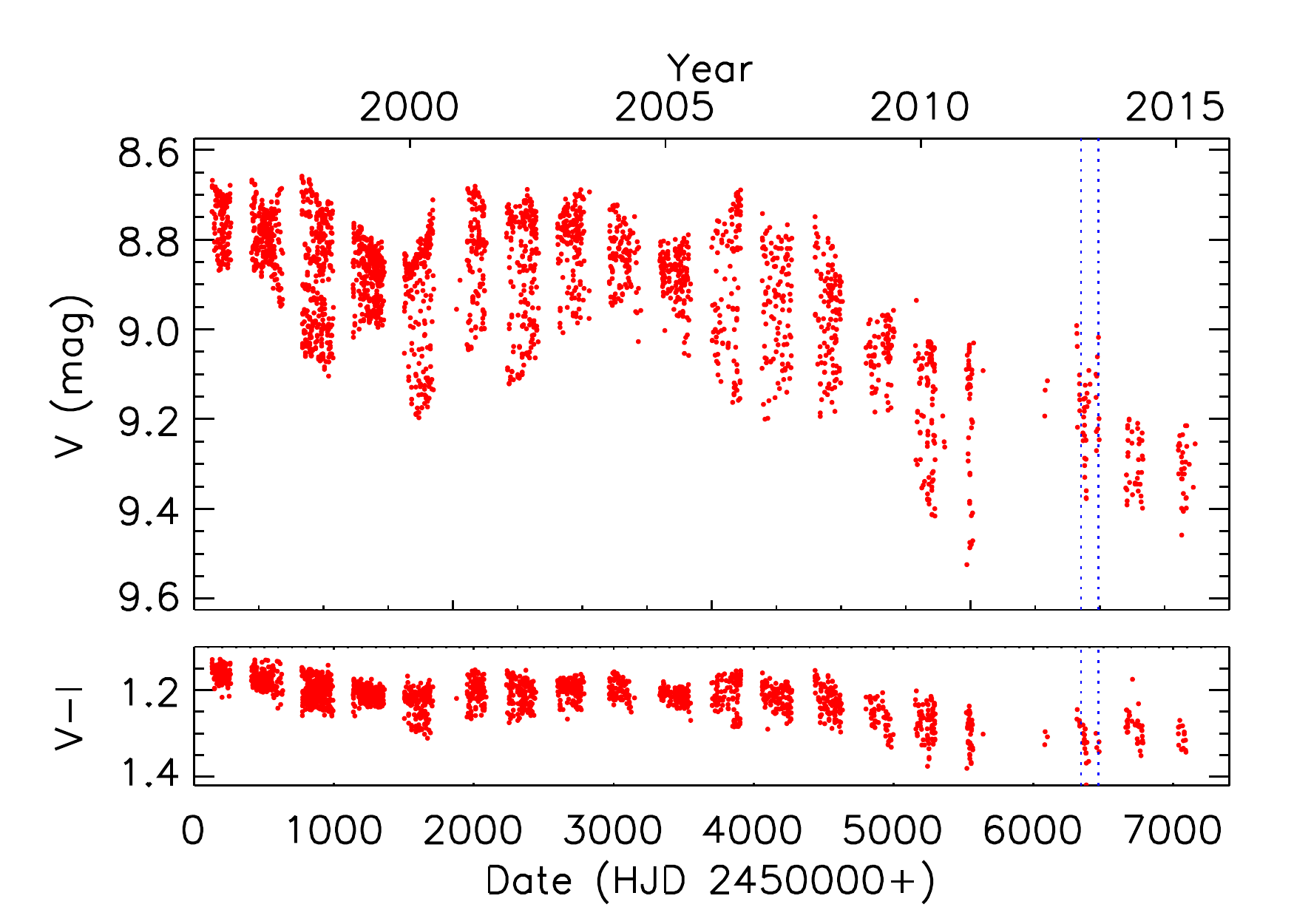}
 {\bf   b)}

  \includegraphics[angle=0,width=87mm,clip]{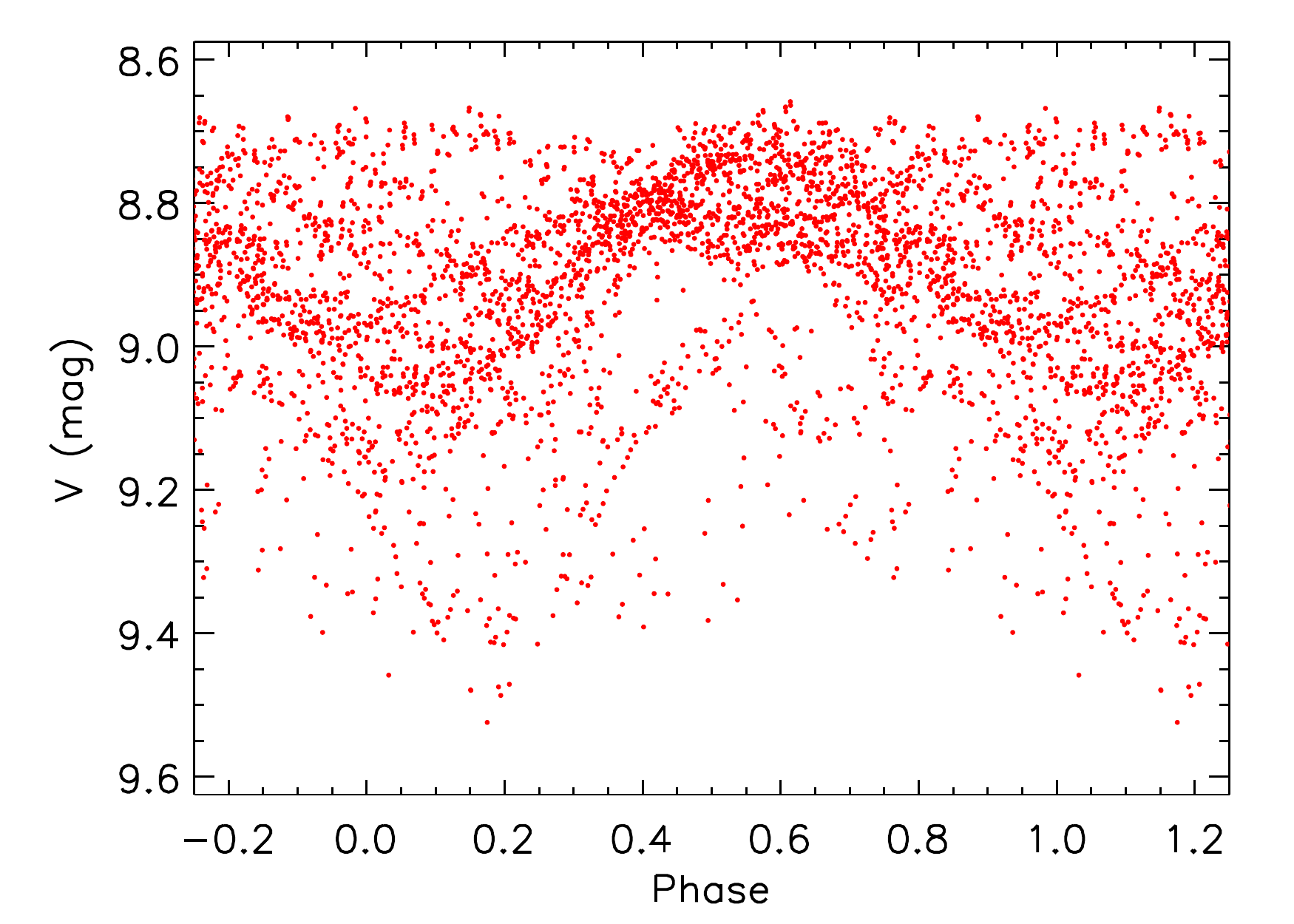}

  \caption{APT photometry of HU\,Vir. $a.$ $V$ brightness (upper panel) and $V-I$ color (lower panel) versus time. The dotted blue lines denote the time range of the STELLA observations in this paper. $b.$ $V$-band light curve of the same data as in panel $a,$ but plotted versus phase from Eq.~\ref{eq:phase}. }
   \label{F2}
\end{figure}

\subsection{Average stellar rotation period and active longitudes}

HU\,Vir represents a classical example for rotationally-modulated light curves. Figure~\ref{F2}a shows our entire APT $V$-band photometry from 1996 to 2015. It nicely indicates the systematic variations over time. A Lomb-Scargle periodogram from its pre-whitened $V$-band data reveals a single, strong peak at 0.0962~c/d, or 10.3909 $\pm$ 0.0080~d. 
Its window function is exceptionally clean for ground-based observations and is due to the continuous (robotic) operations of approximately 20~years. The error of the period is estimated from the full width at half maximum (FWHM) of a  Gaussian fit following \cite{2003drea.book.....B}.
We expect that the rotational signal is subtly affected by the fact that HU\,Vir is differentially rotating because spots may appear at different latitudes and thus with different periods. The observational error is consequently a combination of a measuring errors and an unknown intrinsic variation.

\begin{table}
\begin{flushleft}
\caption{Doppler image log.}
\label{T2}
\begin{tabular}{llllll}
\hline\hline\noalign{\smallskip}
DI & Mean HJD & $\Delta t$ & $N$ & $\phi$-gap & RMSD \\
      & (245+)  & (d)  &  &  &     \\
\noalign{\smallskip}\hline\noalign{\smallskip}
   1  & 6345.56 &  9 & 8  & 0.285 & 0.0024 \\
   2  & 6368.69 & 10 & 17 & 0.184 & 0.0029 \\
   3  & 6380.27 & 10 & 15 & 0.106 & 0.0029\\
   4  & 6392.91 &  9 & 12 & 0.128 & 0.0028 \\
   5  & 6411.08 &  9 & 8  & 0.196 & 0.0029 \\
   6  & 6420.48 &  8 & 9  & 0.233 & 0.0031 \\
   7  & 6435.20 &  9 & 9  & 0.193 & 0.0027\\
   8  & 6446.23 & 10 & 10 & 0.189 & 0.0027 \\
   9  & 6457.72 & 10 & 10 & 0.193 & 0.0027 \\
\noalign{\smallskip}\hline
\end{tabular}
\tablefoot{Successive columns list the mean heliocentric Julian date (HJD) of the image, the time $\Delta t$ elapsed in days, the number of spectra used for the inversion ($N$), the largest phase gap ($\phi$-gap), and the root-mean-square deviation (RMSD) between observed and calculated line profiles.}
\end{flushleft}
\end{table}

Searching the RV jitter for periodicity (after the removal of the orbital variation) is an alternative method to determine the stellar rotation period, although not as precise as from photometry. Subtle RV modulations are common in active stars and are also due to their starspots moving in and out of view, typically with the rotation period of the star if the spot lives long enough. A brief summary of these spot-induced RV detections was given by \cite{2011A&A...535A..98S}. The peak-to-peak amplitude of the RV jitter in HU\,Vir appears to be as large as 4~\kms\ with three pronounced minima over time. This is the largest recorded modulation for any star in the literature so far and hints at a persistent non-axisymmetric spot distribution. We performed an L-S analysis of the STELLA RV residuals and found the best-fit period to be 5.175 $\pm$ 0.010~d. This is almost exactly half of the photometric period and indicative of, on average, a bi-modal spot distribution.

Fig.~\ref{F2}b plots the full data set versus orbital phase. The accumulated light curves hint at the existence of two active longitudes because the star appeared consistently brightest during the times of phase $\approx$ 0.5 and faintest and most variable around phase $\approx$ 1.0, i.e., 180$^\circ$ apart on the stellar surface and with respect to a time of maximum positive RV. This means that the so-called spottedness changes most extensively on the hemisphere that follows (with respect to the orbital motion), but is most stable in the hemisphere that leads, always with respect to the direction of motion of the primary in its orbit around the joint center of mass. A time-dependent and phase-resolved study of this phenomenon on HU Vir and other stars is currently in preparation (Korhonen et al., in prep.).

\subsection{Other stellar parameters}

We employ the spectrum-synthesis method described in \cite{2006ApJ...636..804A} implemented in our PARSES software package \citep{2006ApJ...636..804A,2013POBeo..92..169J}, which itself is included in the SES data reduction pipeline \citep{2008SPIE.7019E..0LW}. It fits synthetic spectra to most \'echelle orders between 480--750~nm using MARCS model atmospheres \citep{2008A&A...486..951G}. Synthetic spectra are first tabulated for a large range of metallicities, surface gravities ($\log g$), temperatures ($T_{\rm eff}$), microturbulences and projected rotational velocities ($v\sin i$) and then fitted iteratively. 
Owing to the large line-profile variations PARSES converged mostly at odd combinations of $\log g$, $T_{\rm eff}$, $v\sin i$, and metallicity, so that we had to rerun the code with fixed $v\sin i$, microturbulence and logarithmic gravity.

The adopted mean values for HU\,Vir are $T_{\rm eff}$ of 5000 $\pm$ 100~K, a gravity $\log g$ of 3.5, a $v\sin i$ of 25 $\pm$ 1~\kms, a microturbulence of 1.5 $\pm$ 0.1~\kms,\ and a solar metallicity. A radial-tangential macroturbulence of 3.0~\kms\ was adopted from \cite{2005oasp.book.....G}. Errors are based on the rms of the entire time-series. We note that the time series of $T_{\rm eff}$ values exhibits a modulation with an average amplitude of 40~K. An L-S periodogram confirms this and finds the strongest peak again at the half of the photometric period of 5.195~d, just like the RV jitter, and the second strongest period at the nominal photometric period of 10.39~d. Furthermore, the phase dispersion minimization method suggested in \cite{1978ApJ...224..953S} found the strongest period at 10.39~d.

The revised  \emph{Hipparcos} parallax of 7.85 $\pm$ 1.14~mas \citep{2007A&A...474..653V} corresponds
to a distance of 127$^{+22}_{-16}$\,pc. A \emph{Gaia} parallax is not yet available. At this distance, the photometry is already affected by interstellar extinction. We adopt a mean extinction value from \cite{2000ApJS..130..201H} for the absolute magnitude calculation ($A_V = 0.8$ mag\,kpc$^{-1}$ and $E(B-V) = A_V/3.3$). The brightest recorded $V$ magnitude of 8\fm66 in Fig.~\ref{F2} then converts  into an absolute visual magnitude of $M_V =3\fm0 \pm 0.3$. Its dereddened $B-V$ color, based on the Tycho color of 0\fm97 $\pm$ 0.01, is 0\fm94. This can be converted to a $T_{\rm eff}$ of $\approx 4960$\,K with the transformation of \cite{1996ApJ...469..355F}.

\begin{figure}
\includegraphics[angle=0,width=87mm,clip]{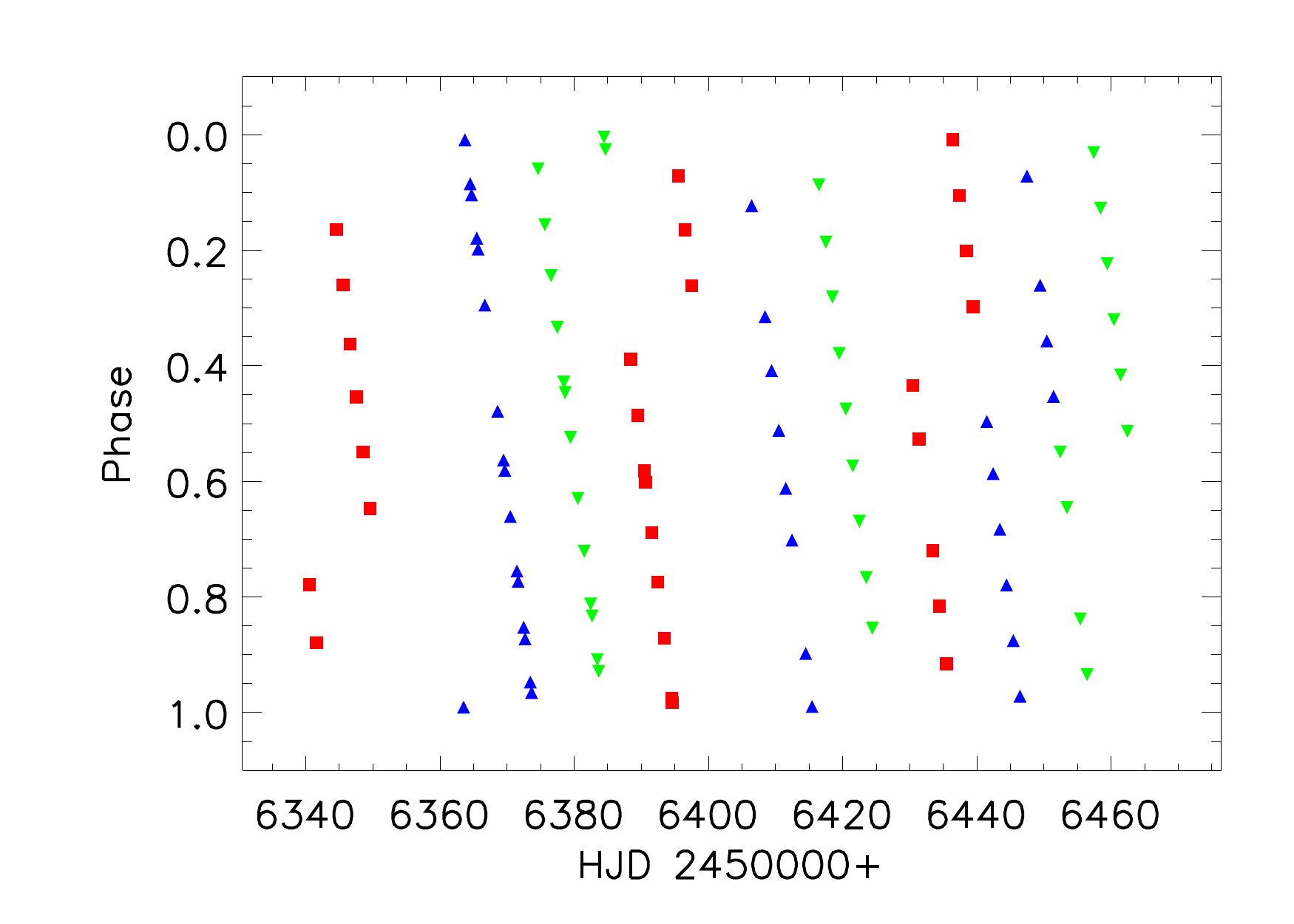}
\caption{Phase coverage of the Doppler images. Different colors and symbols correspond to the phases of individual Doppler images. Phase is computed with the ephemeris in Eq.~\ref{eq:phase}.}
\label{fig:phasecov}
\end{figure}

The bolometric magnitude of HU\,Vir is obtained with a bolometric correction of --0\fm304 \citep{1996ApJ...469..355F} and converts $M_V$ into a luminosity of 6.4 $\pm$ 1.9~L$_\odot$. With an adopted effective temperature of 5000\,K and a bolometric magnitude of the Sun of +4\fm74 the radius of HU\,Vir would then be 3.4 $\pm$ 0.5~R$_\odot$. This is significantly smaller than the minimum radius of 5.1~R$_\odot$ from $P$ and $v\sin i$, which is not uncommon for very active spotted stars and likely due to the inhomogeneous temperature structure.

\section{Doppler imaging}\label{S4}

\subsection{The imaging code iMap}

We employ our DI and ZDI-code \emph{iMap} \citep[e.g.,][]{2012A&A...548A..95C} for computing the Doppler maps. The code performs a simultaneous multi-line inversion of a selected number of photospheric line profiles. For the local line-profile calculation, the code utilizes a radiative transfer solver \citep{2008A&A...488..781C}. The atomic line parameters are taken from the VALD database \citep{1999A&AS..138..119K}. Throughout this paper, we adopted Kurucz ATLAS-9 model atmospheres \citep{2004astro.ph..5087C} which are interpolated for each desired temperature, gravity, and metallicity during the course of the inversion. For all temperature maps in this paper the surface segmentation is set to a $5^\circ\times5^\circ$ partition, resulting in 2 592 surface segments.

We used \emph{iMap} in its latest version which replaced the conjugate gradient method combined with a local entropy regularization \citep{2007AN....328.1043C} with an iteratively regularized Landweber method \citep{2012A&A...548A..95C}. The Landweber iteration rests on the idea of a simple fixed-point iteration derived from minimizing the sum of the squared errors. For details we refer to \cite{2012A&A...548A..95C}.

\subsection{Data preparation and assumptions}

Spectral observations cover the range between Feb 17 and Jun 22, 2013 on a nightly basis, allowing to reconstruct nine consecutive Doppler images with a number of spectra between eight and 17 for each image. Table~\ref{T2} summarizes our Doppler-image log.

The nominal phase resolution between two consecutive spectra is about 0.1 because the observations were carried out on a nightly basis and the rotation period is about  ten days. For images with bad weather or technical issues, the phase gaps were larger than nominal, ranging from between 0.13--0.29 for the largest gaps. A graphical presentation of the phase coverage for each of the Doppler images is shown in Fig.~\ref{fig:phasecov}.

For the line-profile inversion, we used 40 selected absorption lines simultaneously, which were already listed by \cite{2015A&A...578A.101K} in their Table 2. These lines were chosen from the VALD database on the basis of having a minimum line depth of 0.75 $I/I_{C}$, being almost blend-free, and having a good continuum definition around them.

An initial inclination of the rotation axis with respect to the plane of the sky was adopted from \cite{1994A&A...281..395S}. We performed a series of test inversions covering values of inclinations between 30--90$^\circ$ and found the best fit at an inclination of $55 \pm 10^\circ$, which is adopted as the most probable inclination for HU\,Vir. We note that this is different to our previous value of 65\degr\ by 10\degr\ but identical to the value used by \cite{1998A&A...330..541H}. The observed and inverted line profiles of each Doppler image are given in Fig.~\ref{FigA.1}.

\begin{figure}
   \includegraphics[width=\hsize, height=\hsize,keepaspectratio]{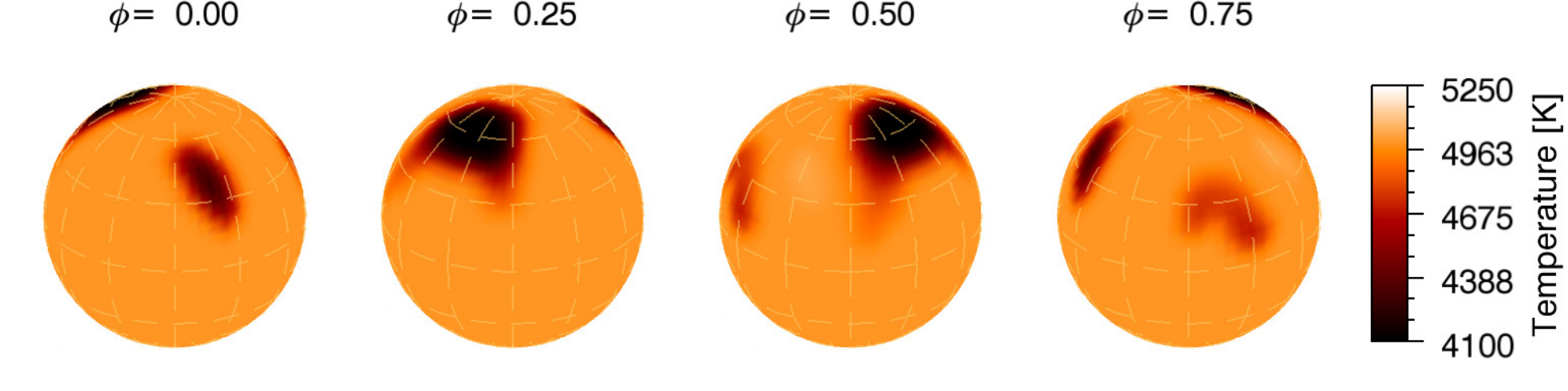}

   \includegraphics[width=\hsize, height=\hsize,keepaspectratio]{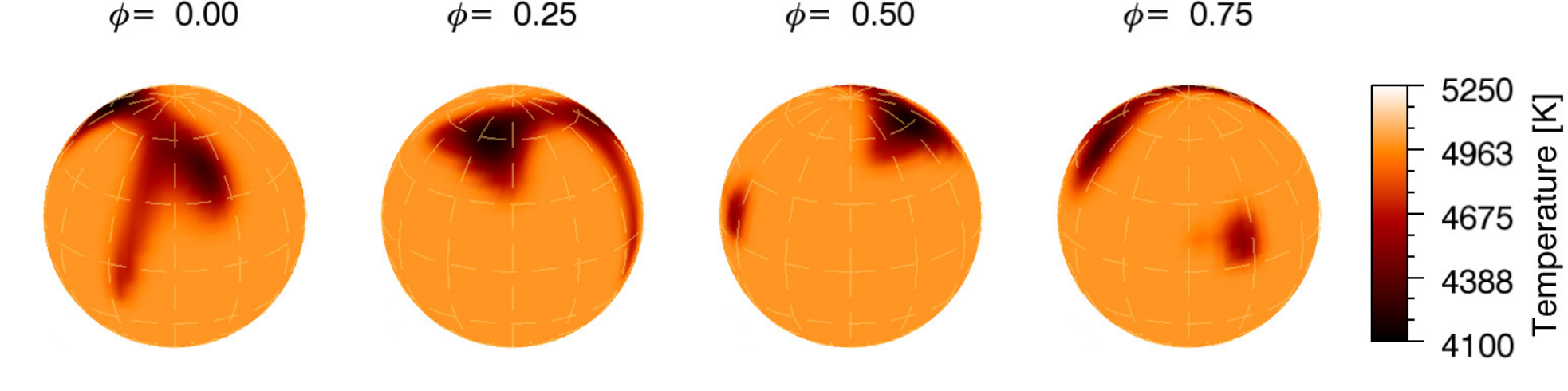}

   \includegraphics[width=\hsize, height=\hsize,keepaspectratio]{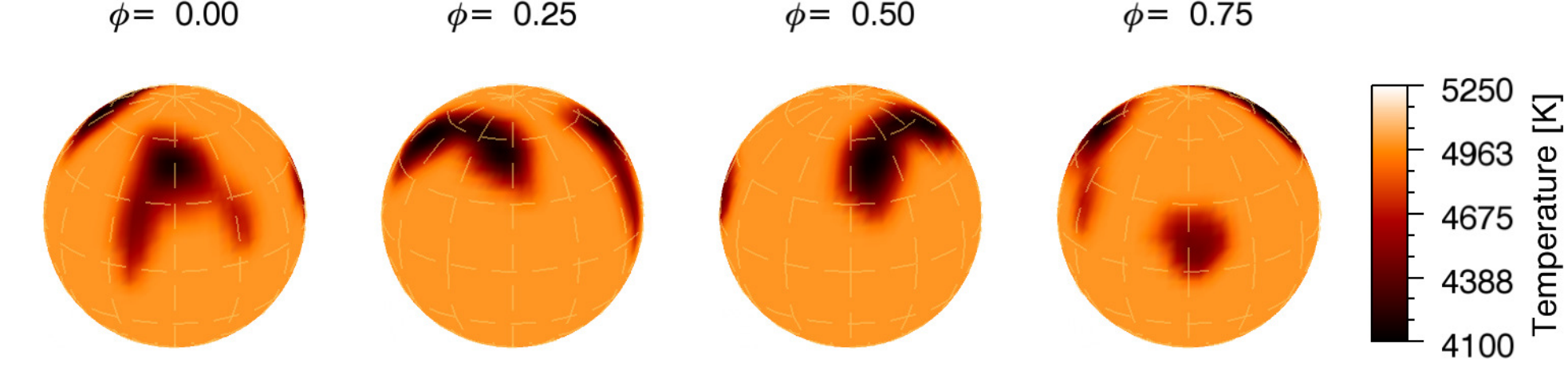}

   \includegraphics[width=\hsize, height=\hsize,keepaspectratio]{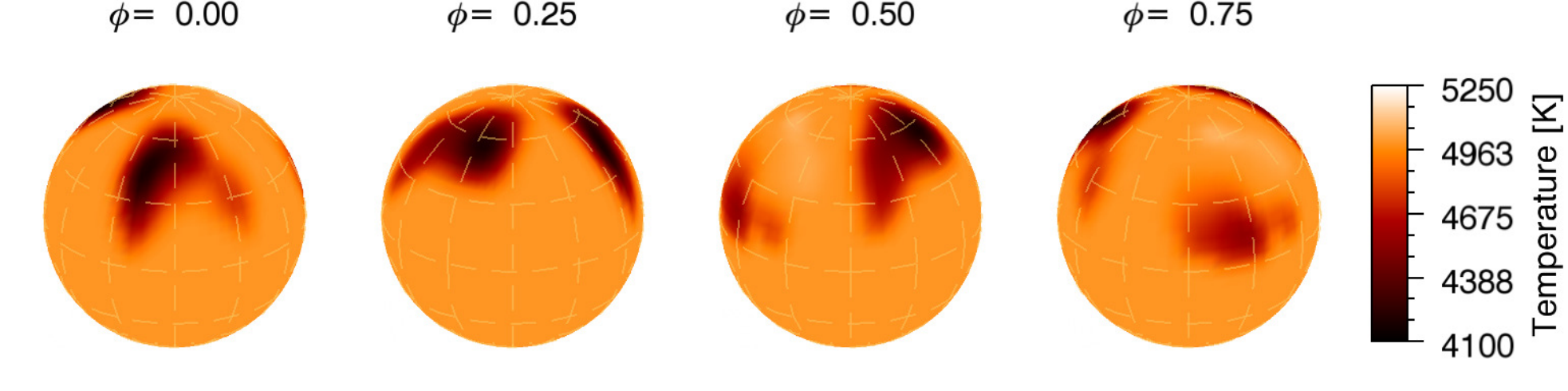}

   \includegraphics[width=\hsize, height=\hsize,keepaspectratio]{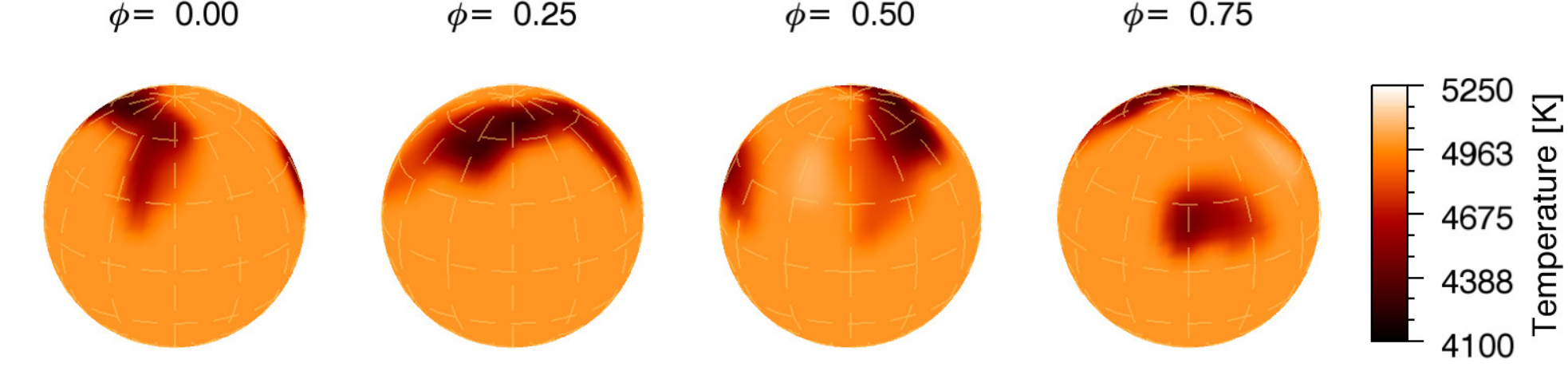}

   \includegraphics[width=\hsize, height=\hsize,keepaspectratio]{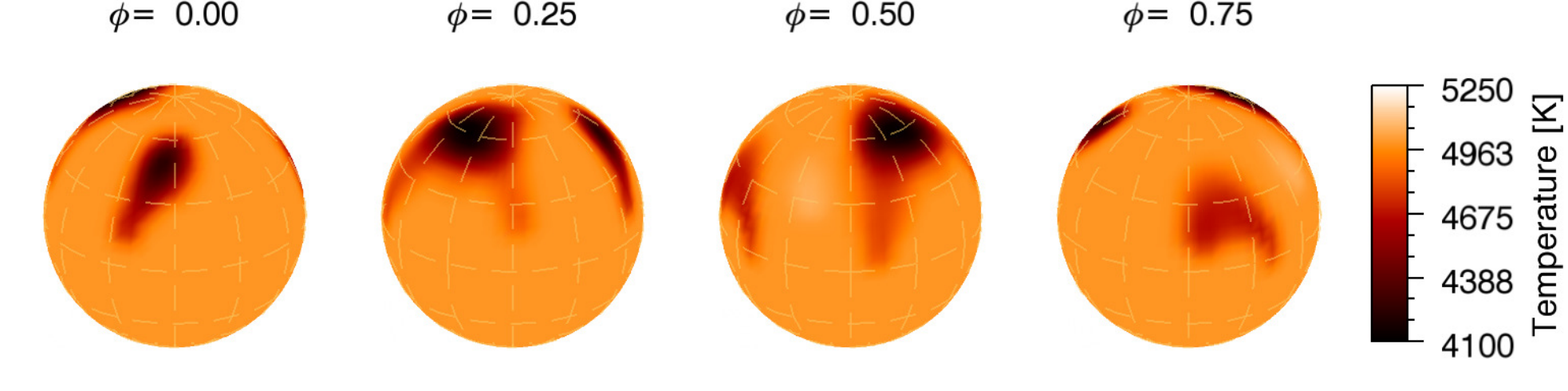}

   \includegraphics[width=\hsize, height=\hsize,keepaspectratio]{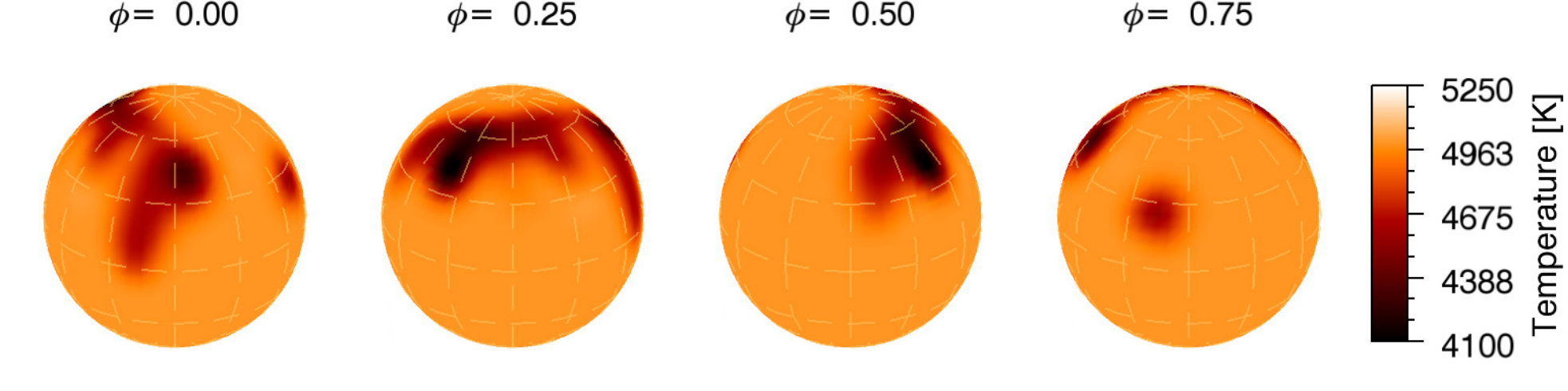}

   \includegraphics[width=\hsize, height=\hsize,keepaspectratio]{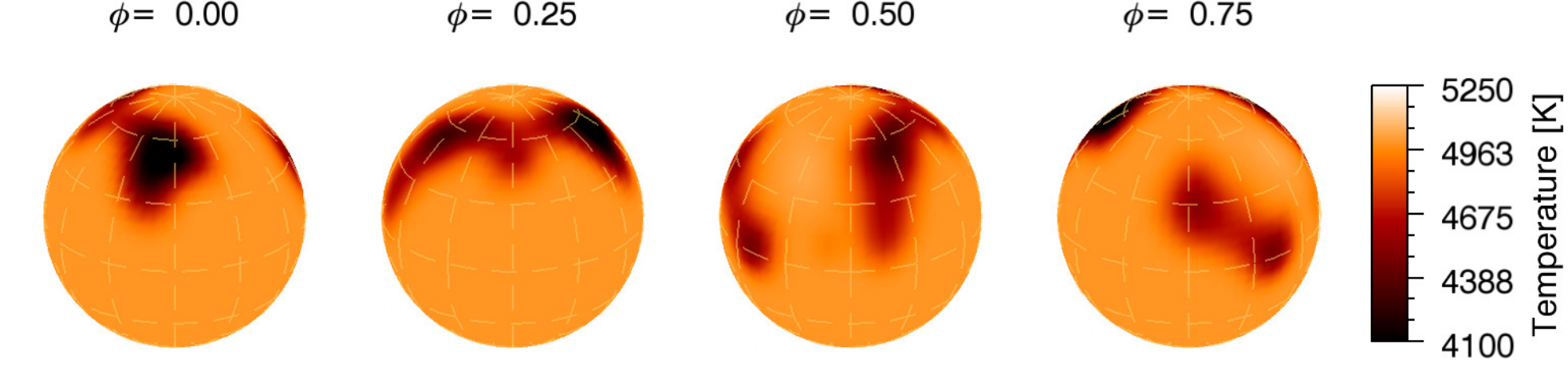}

   \includegraphics[width=\hsize, height=\hsize,keepaspectratio]{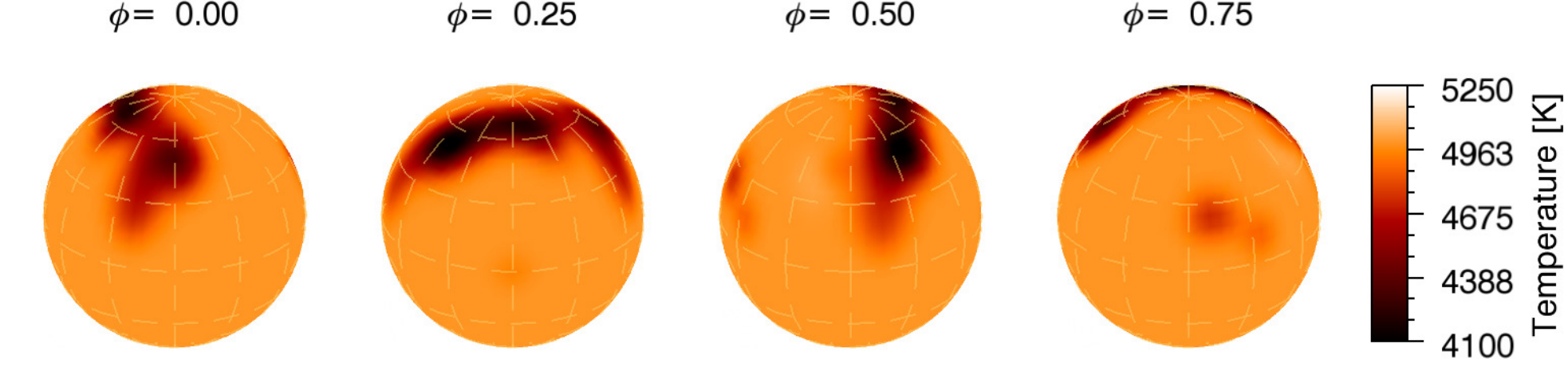}

   \caption{Doppler images of HU~Vir in 2013. The nine images (from top to bottom DI~1 to DI~9) sample 11 subsequent stellar rotations. Maps are shown in four spherical projections separated by $90^{\circ}$.}
         \label{fig:doppler_maps}
\end{figure}

\subsection{Time-series images of HU\,Vir}

Figure~\ref{fig:doppler_maps} presents all nine consecutive Doppler images (DI~1--9) of HU\,Vir. The star does not show a polar cap-like spot as we had recovered initially from data taken in 1991.3 \citep{1994A&A...281..395S} and, although already weaker, in 1995.1 \citep{1998A&A...330..541H}. The maps from 2013 are dominated by two high-latitude active regions with spot appendages occasionally  arching across the visible pole and even down to the stellar equator with temperatures $\approx$ 900~K below the photospheric value. The locations of these spots are consistently recovered from one stellar rotation to the next. We are therefore quite confident about their reality. The recovered shapes, areas, and temperatures change smoothly with time. We emphasize that the time resolution in our movie is still only one stellar rotation.

No unusual warm features higher than 100~K above the photospheric value are recovered. The lowest spot temperature in 2013 is warmer by several hundreds of degree compared to the 1991.3 images, but in agreement with the 1995.1 images. The 1991.3 image is replotted in the Appendix with the same ephemeris as for the STELLA data in this paper.

At this point we emphasize that the photometry shows HU\,Vir's overall brightness has been constantly fading since 2007 (Fig.~\ref{F2}a). The peak $V$ magnitude had dropped by 0\fm5 and $V-I$ reddened by 0\fm1 since then. Obviously, the star underwent a dramatic change between the time of the two earlier  Doppler-imaging epochs and 2013. We also note that by 2015/16 (the time of writing this paper) the $V$ brightness had dropped yet another 0\fm1 and reached the lowest ever average magnitude of 9\fm3 (compared to the adopted unspotted brightness of 8\fm57 back in 1991 when HU\,Vir was brightest).

This leaves us with the situation that the long-term Doppler images revealed a fading of the polar spot along with a general warming of the spotted regions, while the light and color curves showed a dimming of the star along with a reddening. This would be a contradiction if only cool spots were the activity tracers because a dimming paralleled by a reddening indicates a growing and cooling of the spots (e.g., \citealt{1985ApJ...289..644P}, \citealt{2009A&ARv..17..251S}, and references therein), just the opposite as observed. However, this type of behavior is in agreement with solar observations where, in the course of the sunspot cycle, the brightness contribution due to warm plages and faculae is larger than the dimming due to cool spots. We note that on HU\,Vir and other stars we only resolve a smoothed large-scale spot distribution.

Sinusoidal variations of the H$\alpha$ emission with an amplitude of significantly less than the surface rotation rate of 25\,\kms\ were interpreted as arising from a plage at high latitude and adjacent to the polar appendage seen in the photospheric spot distribution \citep{1994A&A...281..395S, 1998A&A...330..541H}. Therefore, there is evidence that HU\,Vir showed enhanced chromospheric plage activity at earlier dates, at least during 1991.3 and 1995.1, but none in 2013.

\begin{table}
\centering
\caption{Differential-rotation parameters from the best fits with Eqs.~\ref{eq:diffrot1} and~\ref{eq:diffrot2}.}
\begin{tabular}{l | c || l | c }
\hline\hline\noalign{\smallskip}
\multicolumn{2}{c||}{DR, Eq.~\ref{eq:diffrot1}} & \multicolumn{2}{c}{DR, Eq.~\ref{eq:diffrot2}}\\
\noalign{\smallskip}\hline\noalign{\smallskip}
  $ \alpha$   &  --0.029$\pm$0.005 & $\alpha$    & --0.026$\pm$0.009\\
  lap-time (d) & $\approx$360       & lap-time (d) & $\approx$400  \\
  $\Omega_{\rm equ}$ $(^{\circ}/d)$ & 34.38$\pm$0.15 & $ \Omega_{\rm equ}$ $(^{\circ}/d)$ & 34.83$\pm$0.12 \\
  $\Delta\Omega$ $(^{\circ}/d)$ & --1.01$\pm$0.23 & $\Omega_{1} $ $(^{\circ}/d)$ & --2.07$\pm$0.58 \\
  & & $\Omega_{2}  $ $(^{\circ}/d)$& 2.98$\pm$0.55\\

\noalign{\smallskip}\hline
\end{tabular}
\label{table:diffrotpar}
\end{table}

\section{Differential rotation}\label{S5}

\subsection{Technique}

Cross-correlations of Doppler images were previously used to measure DR coefficients for several stars \citep[e.g.,][]{1997MNRAS.291..658D, 2015A&A...573A..98K, Kovari2016, 2015A&A...578A.101K, Ozdarcan2016}. It is only necessary to  take two Doppler images and cross correlate the flux (temperature) along longitude in each latitude band. The method is very powerful if the spot configuration appears structured and has many isolated spots or polar appendages but completely fails, for example, in the case of a single axi-symmetric homogeneous polar spot. The method also fails if the average spot life time is shorter than the time between the Doppler images because, then, the cross-correlation would pick up an artificial signal owing to spot formation or decay rather than DR. It is thus essential that the Doppler images properly sample the spot migration in time.

We apply this method to our nine Doppler images from 11 consecutive stellar rotations obtaining eight cross-correlation-function (CCF) maps. This data situation is ideal because, firstly, spots can be redundantly identified and, secondly, the time range is long enough that significant spot migration can take place. Furthermore, we can average our eight CCF maps to increase their validity.

To find the best correlation, Gaussians are fitted to the CCF maps per latitude bins of $5^{\circ}$ and their maximum referred to as the best-fit longitudinal shift. The error bars represent the FWHMs of the Gaussians. 
We note that the FWHMs increase near the pole as a result of Mercator projection.

For the representation of the latitude-dependent cross-correlation pattern, we fit two DR laws by performing non-linear least-squares fits. The most common law used to describe DR on stars is given by
\begin{equation} \label{eq:diffrot1}
\Omega(\theta)=\Omega_{\rm equ} - \Delta\Omega sin^{2}(\theta) ,
\end{equation}
where $\Omega(\theta)$ is the angular velocity at latitude $\theta$, while the surface shear parameter, $\alpha$, is defined as $\Delta\Omega / \Omega_{\rm equ}$, where $\Delta\Omega = \Omega_{\rm equ} - \Omega_{\rm pole}$ is the normalized difference between the angular velocities at the equator and at the pole, respectively.

Another representation of the DR pattern is given by a two-term expression, which is usually used in the solar case
\begin{equation} \label{eq:diffrot2}
\Omega(\theta)=\Omega_{\rm equ} + \Omega_{1} sin^{2}(\theta) + \Omega_{2} sin^{4}(\theta) ,
\end{equation}
where the angular velocity at the pole is now defined as $\Omega_{\rm pole} = \Omega_{\rm equ} + \Omega_{1} + \Omega_{2}$. Consequently, in this case, the shear parameter, $\alpha$, is defined as $-(\Omega_{1} + \Omega_{2})/\Omega_{\rm equ}$.

\subsection{Results}

The final (unweighted) average CCF map is shown in Fig.~\ref{fig:diff_rot}a. The color code is the correlation coefficient in arbitrary units as indicated. The subtle asymmetry of the distribution towards higher latitudes indicates the amount of surface DR.

Fig.~\ref{fig:diff_rot}b quantifies the longitudinal shifts as a function of latitude and shows the fits from both DR laws from Eq.~\ref{eq:diffrot1} and~\ref{eq:diffrot2}, respectively. The best fit parameters are listed in Table~\ref{table:diffrotpar}. Because most spotted regions on HU\,Vir appear at high latitudes, the DR law of Eq.~\ref{eq:diffrot2} with its sin$^4$ $\theta$ term leads to a significantly better representation of the observed shear. We therefore adopt its result of a surface shear of -0.026 $\pm$ 0.009 with a lap time (the time the equator needs to lap the pole) of $\approx$ 400\,d as the main result in this paper. We note that there is only a subtle difference in shear between both interpretations, which strengthens the reliability of our result. Therefore, the detected surface shear is roughly eight times weaker than that of the Sun, and is of opposite sign.

\begin{figure}
{\bf a)}

   \includegraphics[width=100mm, height=\hsize,keepaspectratio]{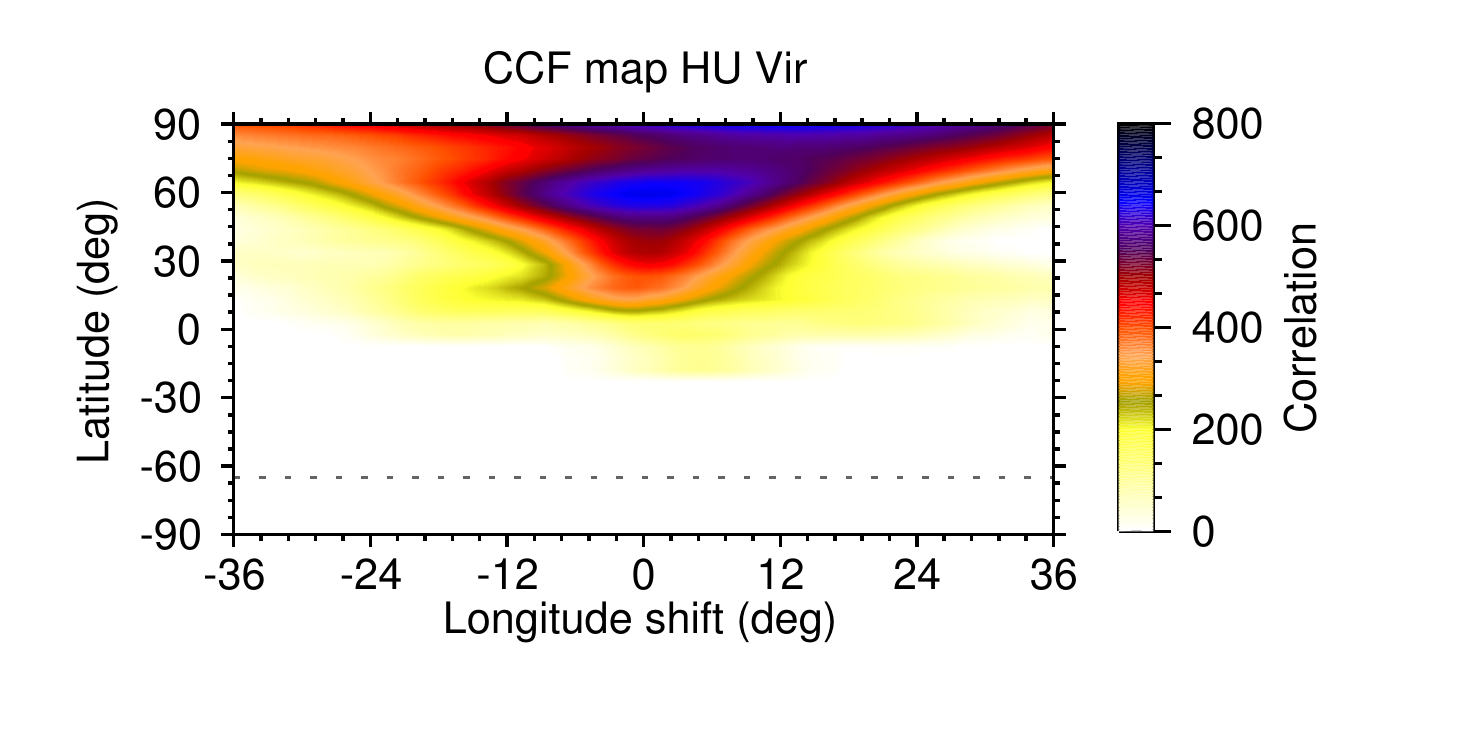}
{\bf b)}

   \includegraphics[width=87mm, height=\hsize,keepaspectratio]{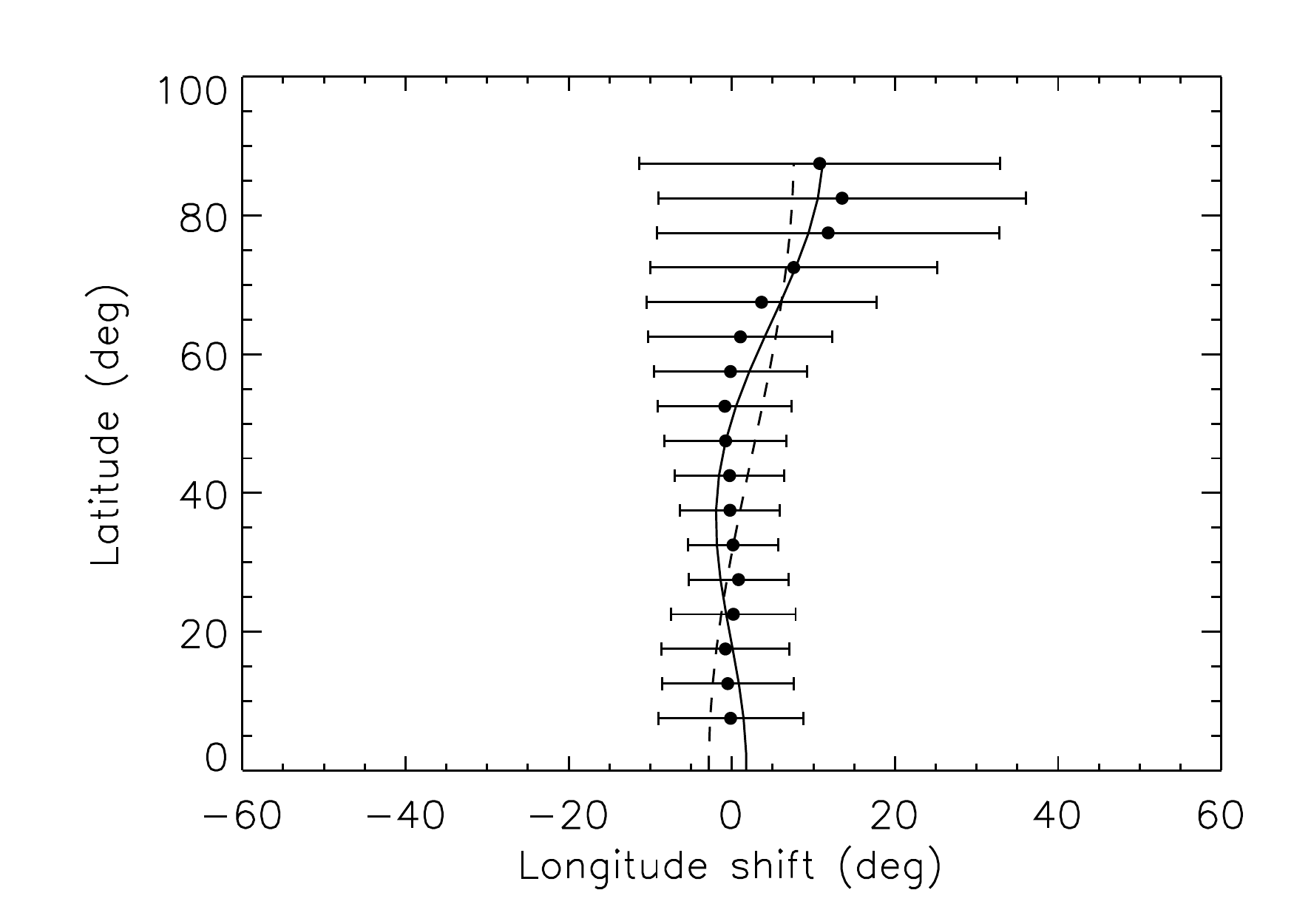}
\caption{Differential rotation signature. $a.$ Average cross-correlation-function map. The darkness
of the shade (color) scales with correlation (the darker the better the correlation). $b.$
Correlation fits per latitude. Dots are the correlation peaks per $5^{\circ}$-latitude bin. The
error bars are defined as the FWHM of the corresponding Gaussian fits. The dashed line represents
the DR-fit using Eq.~\ref{eq:diffrot1} and the solid line is the DR-fit using
Eq.~\ref{eq:diffrot2}.}
   \label{fig:diff_rot}
\end{figure}

\section{Summary and conclusions}\label{S6}

HU\,Vir appears to show large-scale surface activity that is related to its orbital motion. Our 20-year photometric data base tentatively indicates that the following stellar hemisphere, with respect to the direction of the orbital motion, undergoes more spot variability than the leading hemisphere. This can be dubbed a stable active longitude because we have not seen major flips and flops between the 180\degr\ locations in the previous 20 years. Flip flops between two active longitudes are otherwise common among active spotted stars \citep{2007IAUS..240..453K, 2015arXiv151106116K, 2015A&A...578A.101K}, sometimes even with a significant periodicity \citep{Ozdarcan2016}.

Differential rotation of HU\,Vir was previously measured by two independent authors;  \cite{1994A&A...281..395S} derived an anti-solar DR with a shear parameter $\alpha$ of $-0.0228 $, and \cite{1998A&A...330..541H} measured a very similar shear of $\alpha$ of $-0.0205$ and of same sign. Both values were obtained by a direct comparison of spot locations from two Doppler images separated in time, which were dubbed the spot-migration technique. Our new DR measurement employs nine Doppler images from eleven consecutive stellar rotations and is clearly superior to these previous determinations. Nevertheless, it confirms the previously claimed  anti-solar DR and nails its shear parameter to $-0.026 \pm 0.009$ with a lap time to $\approx$ 400\,d; this is in good agreement with the previous results. As such, it  strengthens the reliability of such detections of anti-solar differential rotation.

\begin{acknowledgements}We are grateful to the State of Brandenburg and the German Federal Ministry for Education and Research (BMBF) for their continuous support of the STELLA and APT activities. The STELLA facility is a collaboration of the AIP with the IAC in Tenerife, Spain. We thank Thomas Granzer for his help with the APT data management. Discussions with Carsten Denker and Zsolt K\H{o}v\'ari on various differential rotation issues are kindly acknowledged. We thank the referee, Jaan Pelt, for his valuable comments and suggestions that helped to improve this manuscript. Finally, we thank the Leibniz-Association and its SAW program for supporting one of us (G.H.) through a graduate school grant.

\end{acknowledgements}

\bibliographystyle{aa}             
\bibliography{references}          

\begin{appendix}

\section{Line profiles used for Doppler images}

Fig.~\ref{FigA.1} shows the observed and inverted line profiles for each of nine Doppler images (DI1-DI9). The corresponding RMSDs are given in Table~\ref{T2}.

\begin{figure*}[!htb]
    \includegraphics[clip,trim=15 0 35 0, width=35mm]{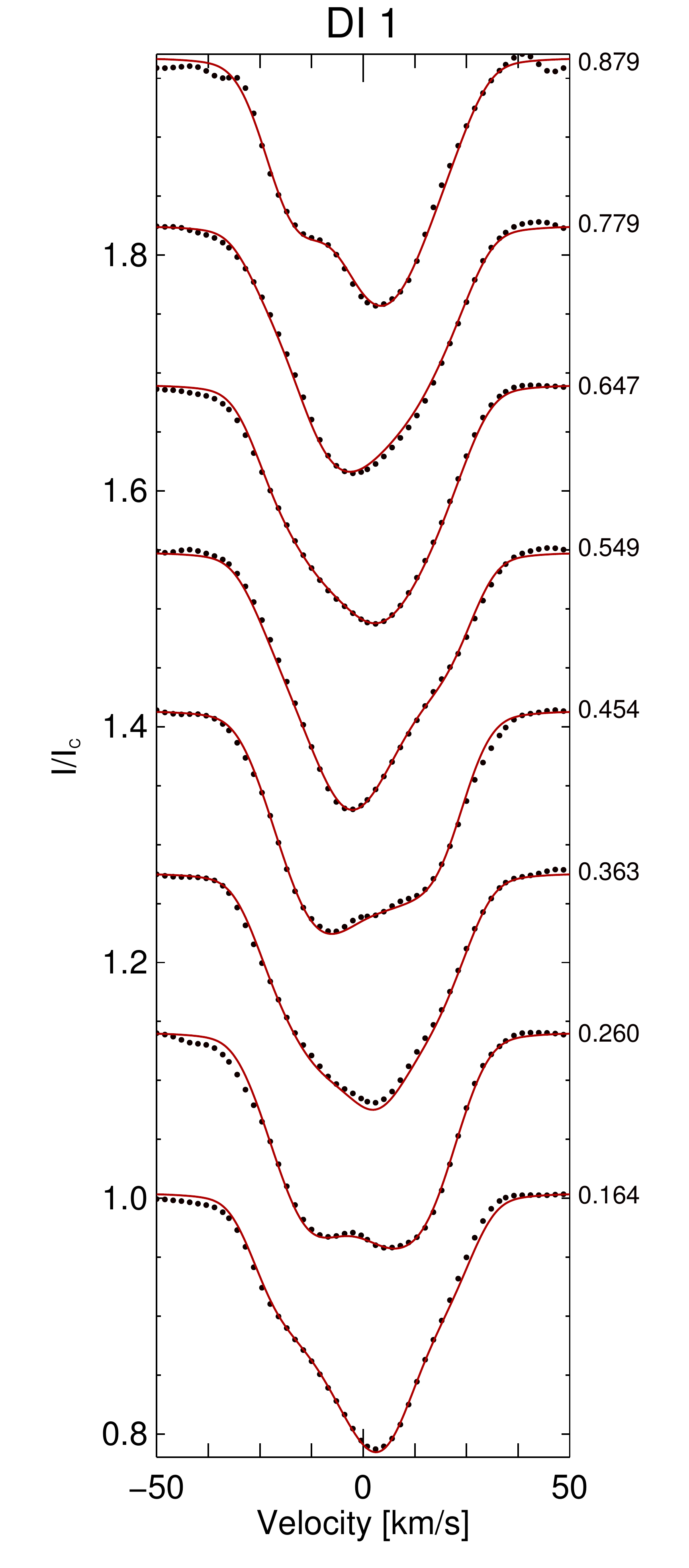}
    \includegraphics[clip,trim=15 0 35 0,width=35mm]{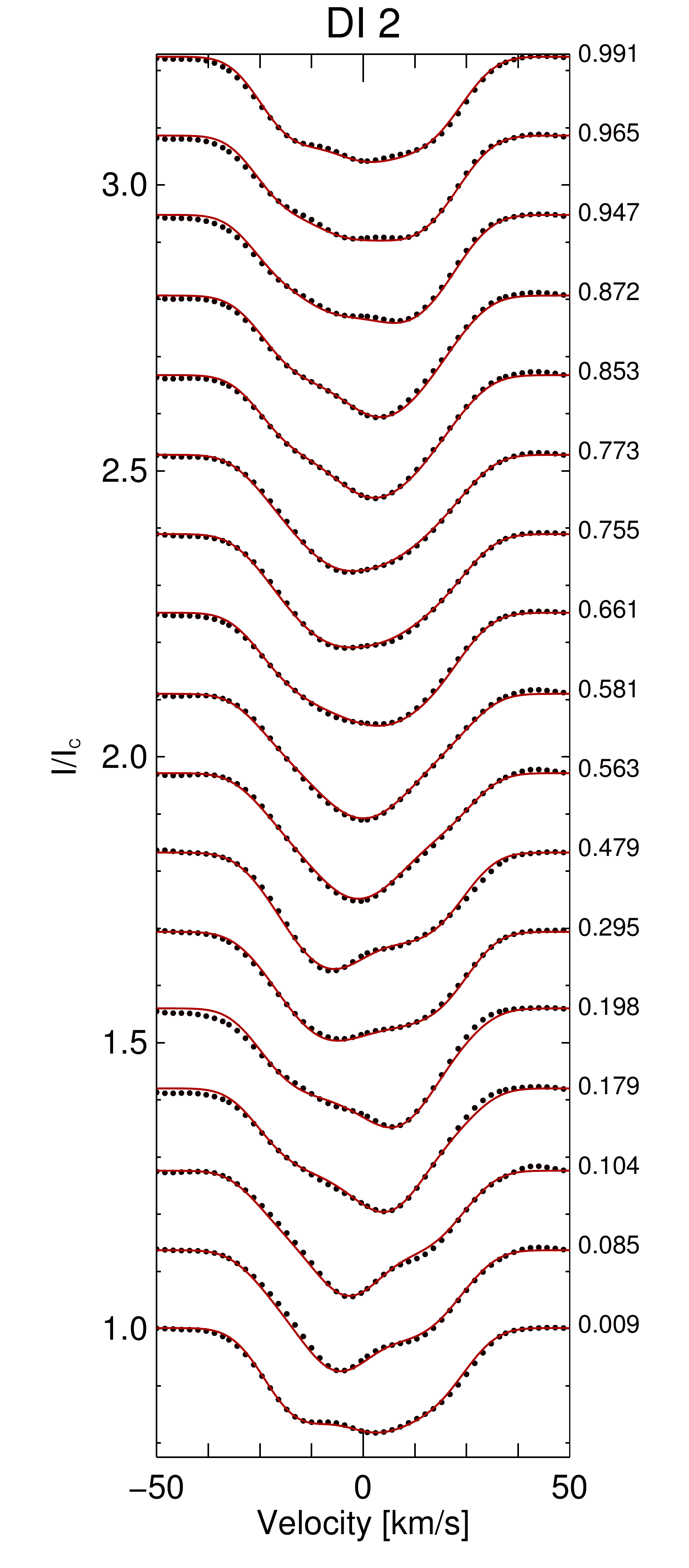}
    \includegraphics[clip,trim=15 0 35 0,width=35mm]{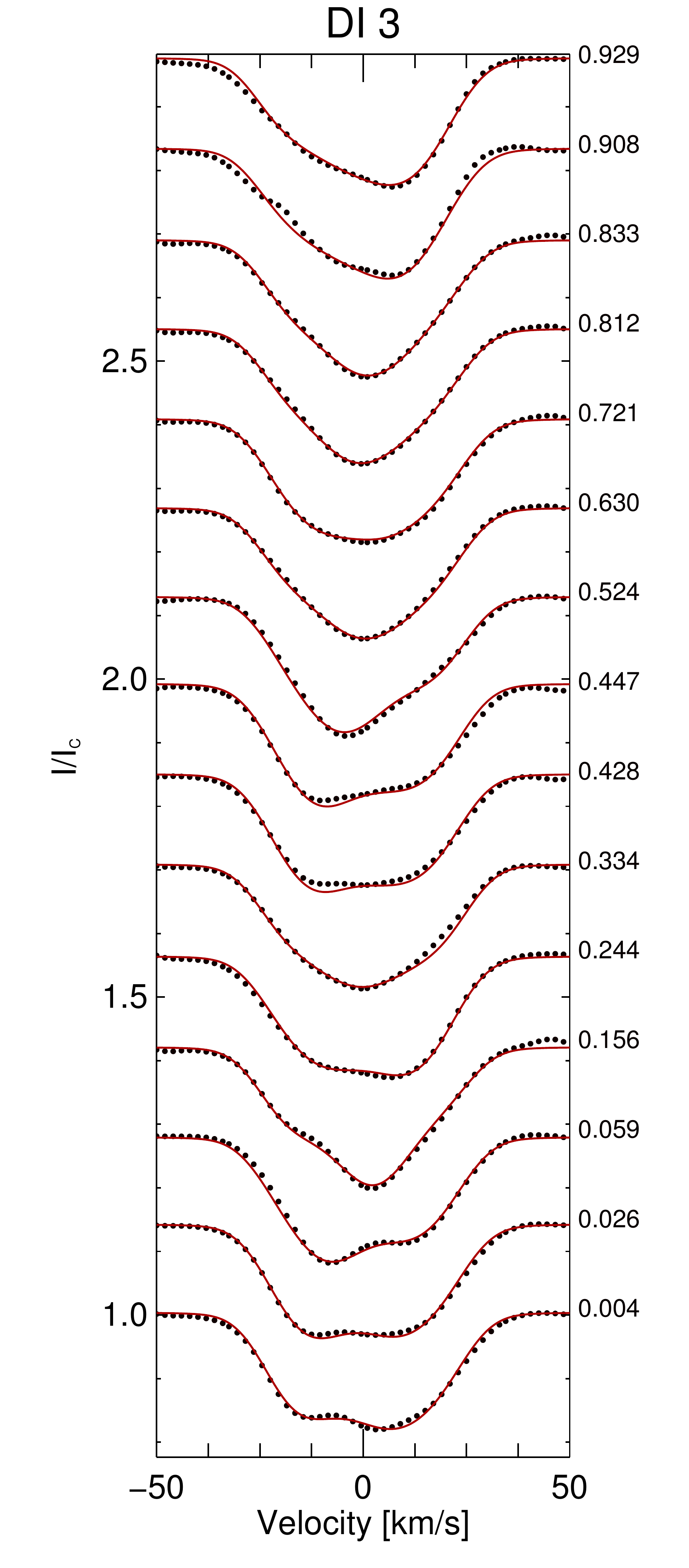}
    \includegraphics[clip,trim=15 0 35 0,width=35mm]{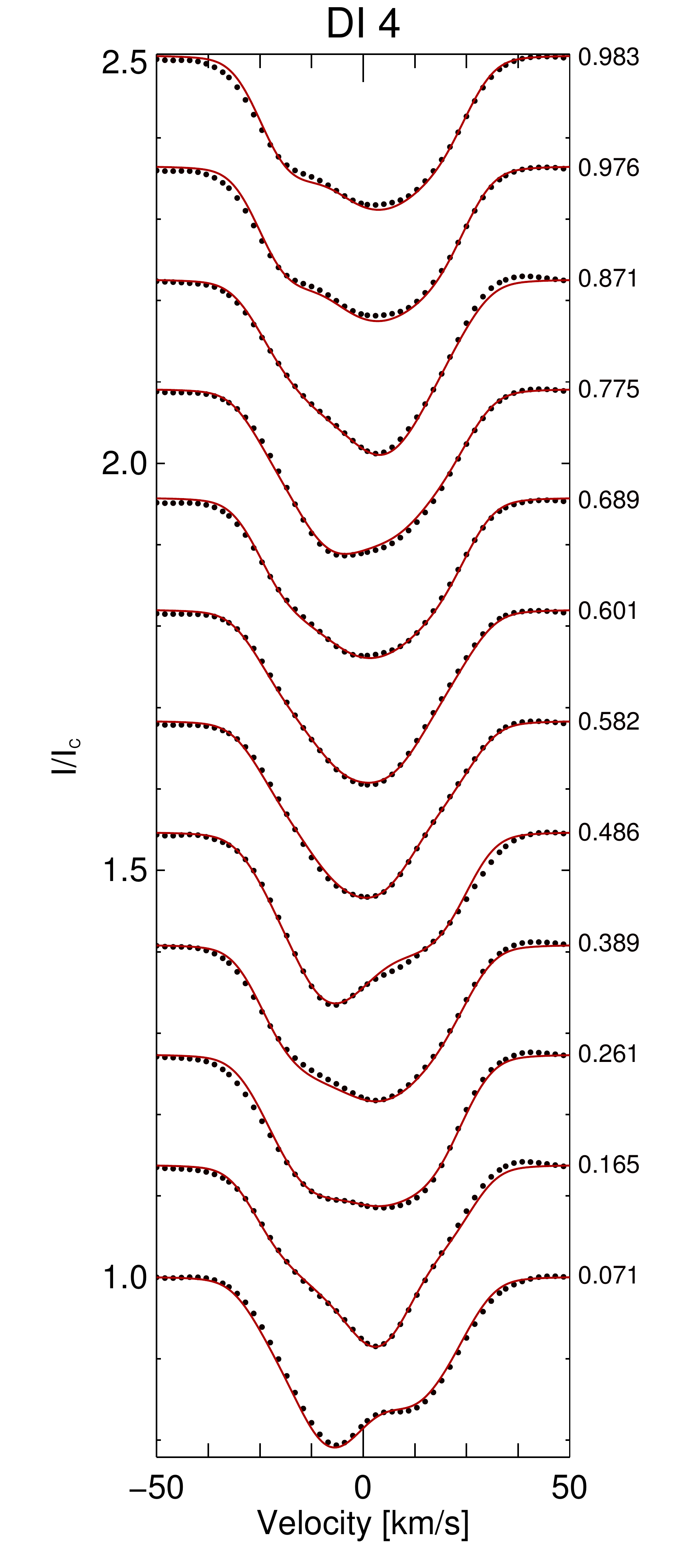}
    \includegraphics[clip,trim=15 0 35 0,width=35mm]{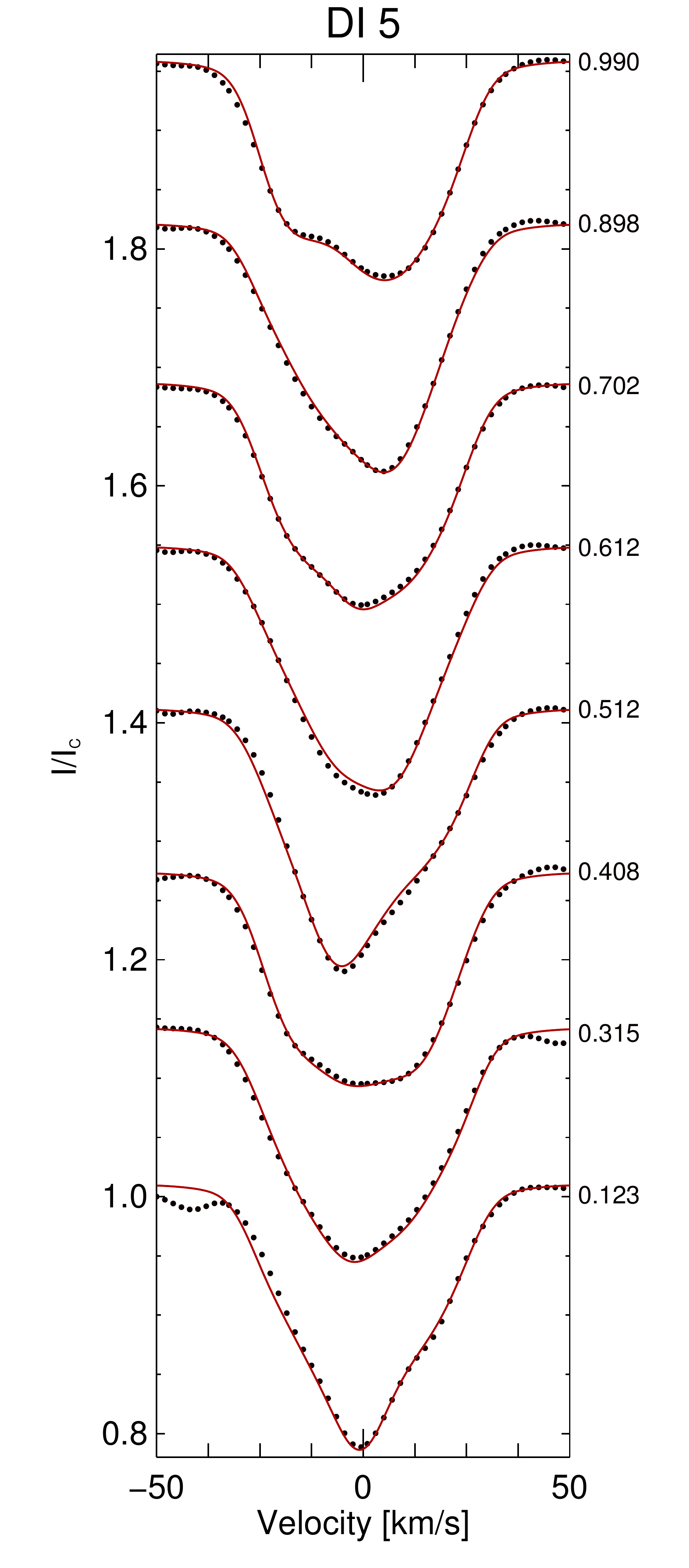}
    \includegraphics[clip,trim=15 0 35 0,width=35mm]{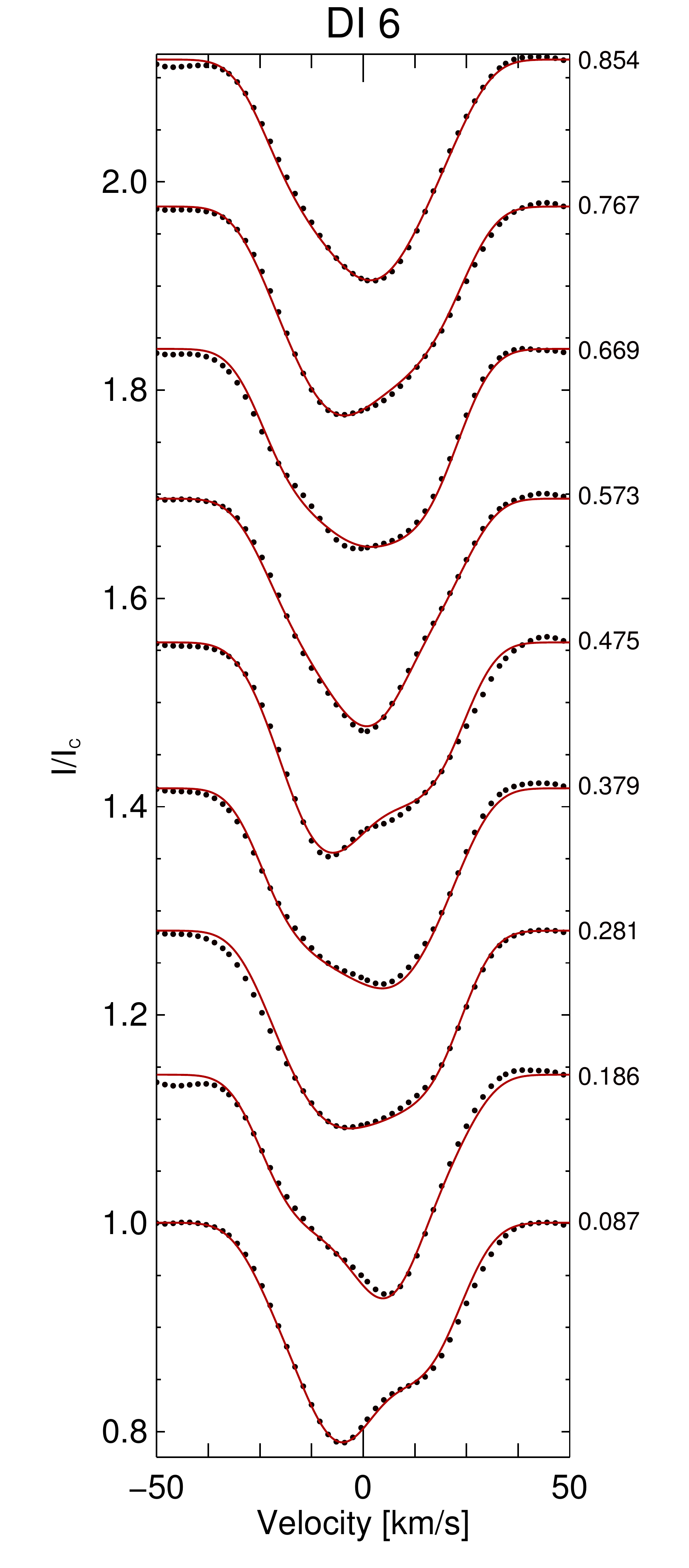}
    \includegraphics[clip,trim=15 0 35 0,width=35mm]{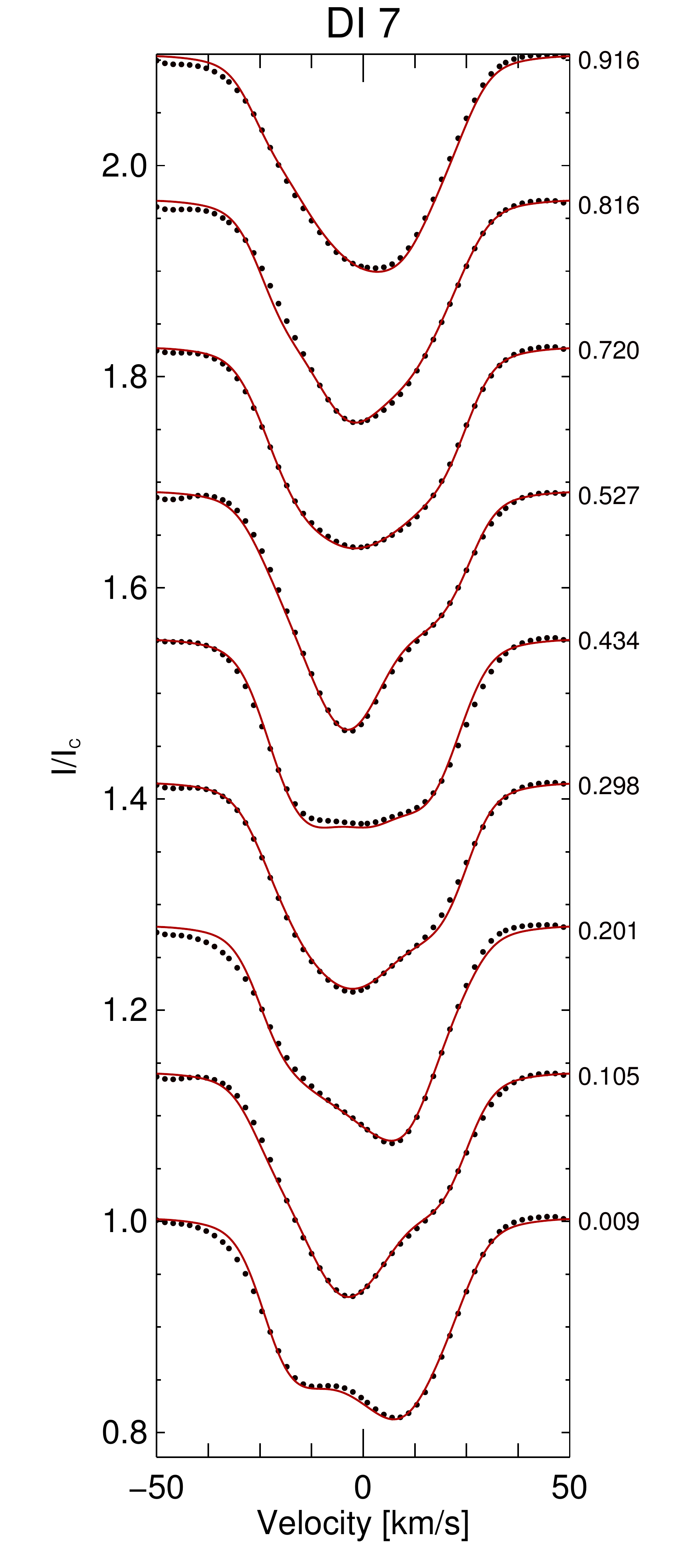}
    \includegraphics[clip,trim=15 0 35 0,width=35mm]{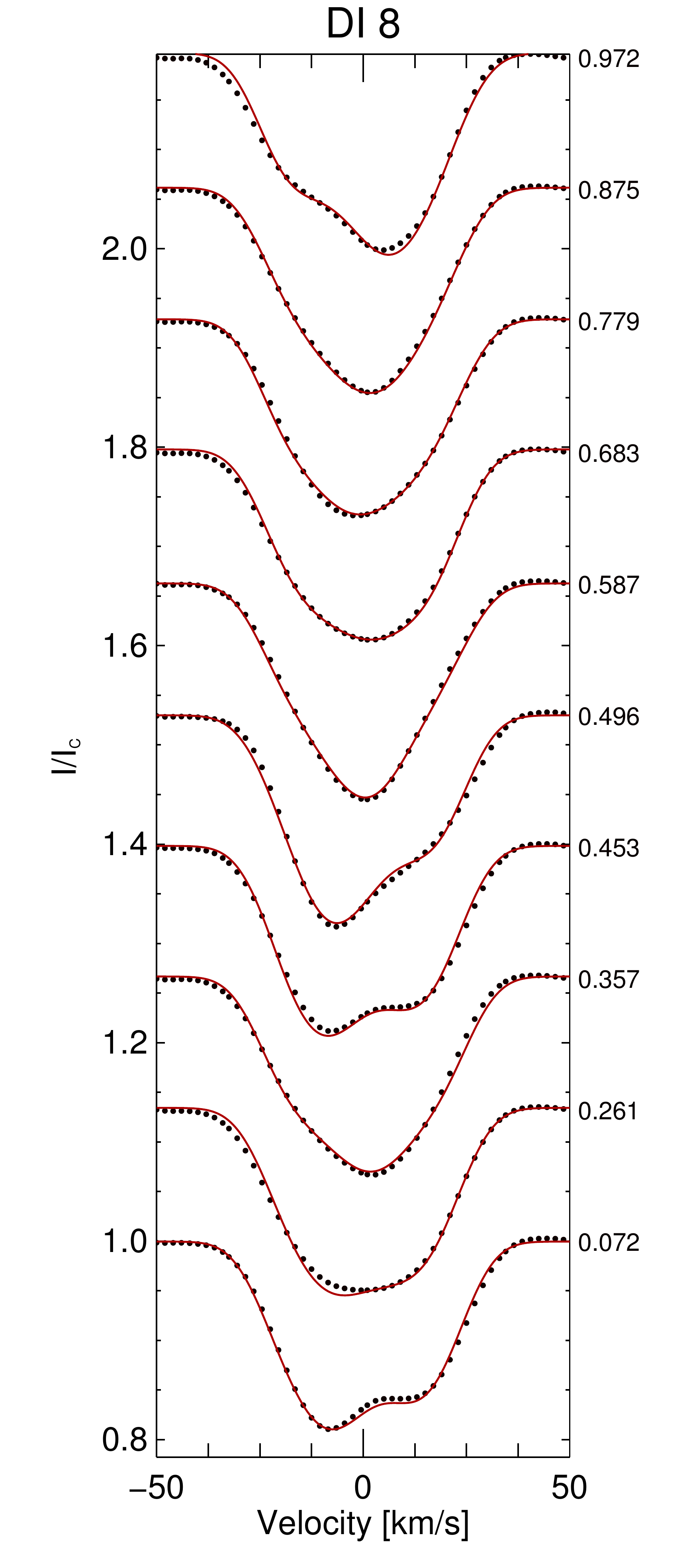}
    \includegraphics[clip,trim=15 0 35 0,width=35mm]{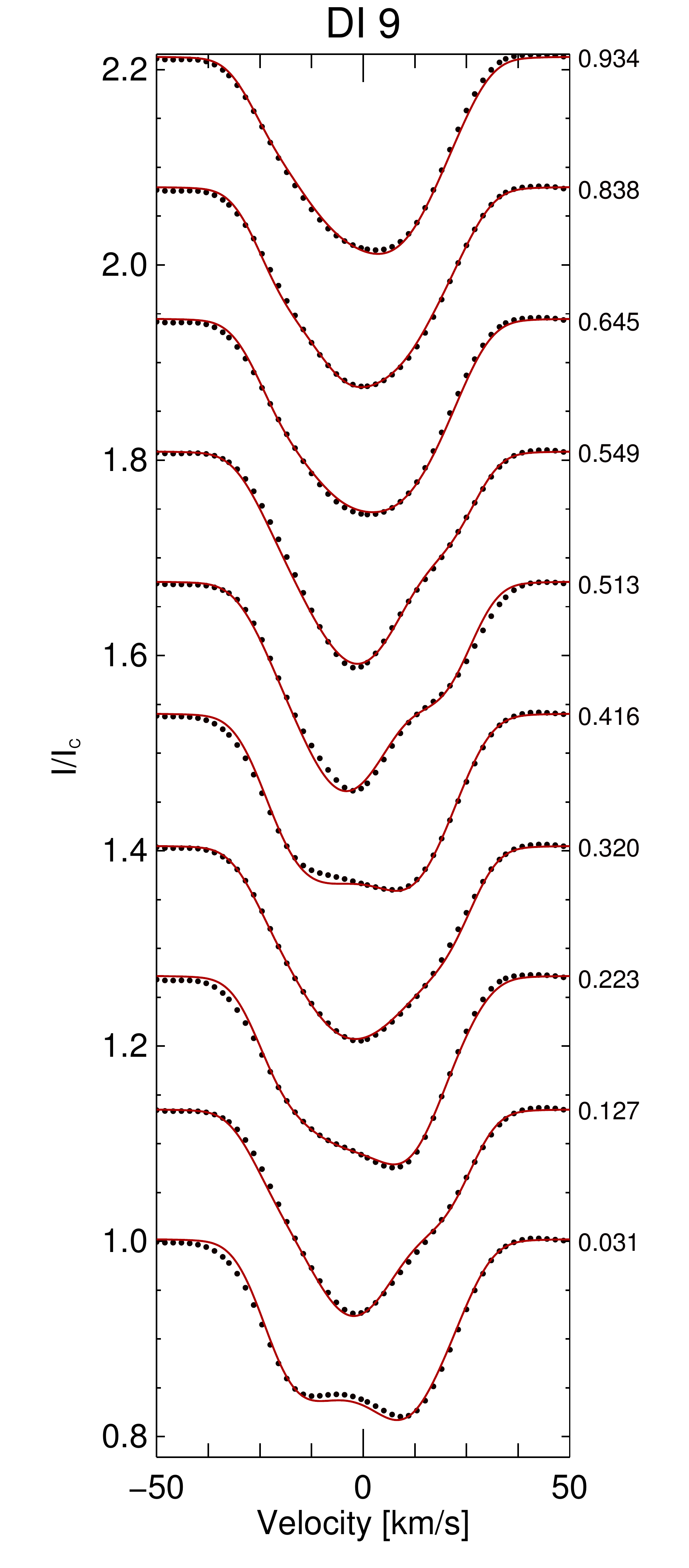}

\caption{Line profiles used for nine Doppler images. Each figure shows the observed (dotted lines) and inverted  (solid red lines) profiles for one Doppler image, stating the number of the image and respective phases. Rotation advances from bottom to  top.}
 \label{FigA.1}
\end{figure*}

\section{Doppler image from 1991.3}

Fig.~\ref{FigB.1} shows the Doppler image from \cite{1994A&A...281..395S}. This is an average from four individual Doppler maps, using one photospheric line each. It shows a large cool polar spot  with two large appendages. The polar spot is cooler by 1 100--1 500~K than the undisturbed photosphere (5 000K).


\begin{figure}[H]
\includegraphics[width=\hsize, height=\hsize,keepaspectratio]{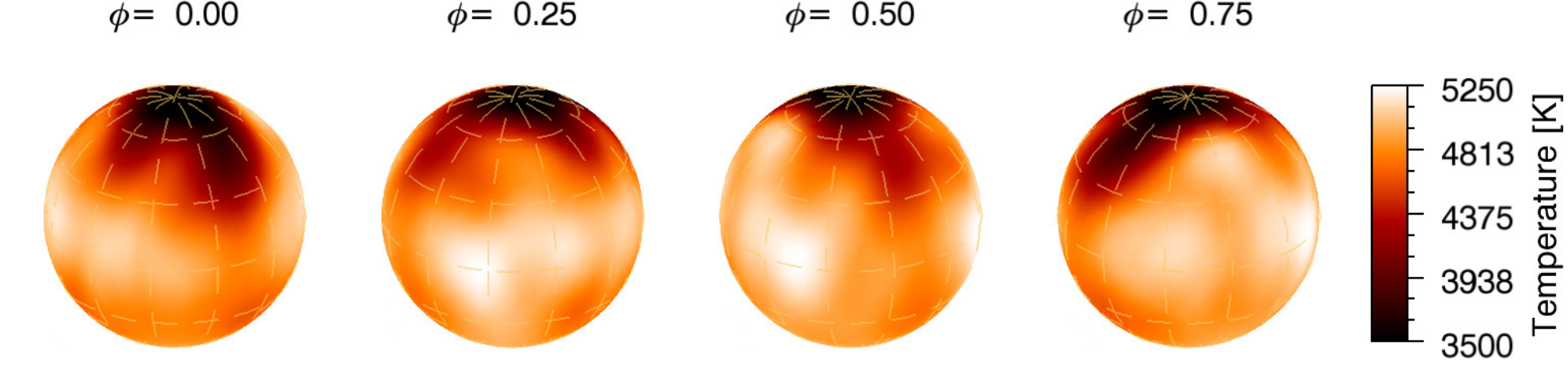}
\caption{Doppler map of HU~Vir from \cite{1994A&A...281..395S}, but with the same ephemeris as for the STELLA data in this paper.}
    \label{FigB.1}
\end{figure}

\end{appendix}


\end{document}